\newif\iffigs\figstrue

\iffigs
   \documentstyle[12pt,epsf]{article}
\else
   \documentstyle[12pt]{article}
\fi

\def\thefootnote{\alph{footnote}}
\def\thefootnote{\fnsymbol{footnote}}
\def\be{\begin{equation}}
\def\ee{\end{equation}}
\def\ba{\begin{eqnarray}}
\def\ea{\end{eqnarray}}
\def\de{\partial}
\def\Br{\langle}
\def\kt{\rangle}
\newcommand{\bea}{\begin{eqnarray}}
\newcommand{\eea}{\end{eqnarray}}

\newcommand{\eq}{\begin{equation}}
\newcommand{\en}{\end{equation}}
\newcommand{\eqa}{\begin{eqnarray}}
\newcommand{\ena}{\end{eqnarray}}

\newcommand{\ZZ}{\hbox{{\rm Z{\hbox to 3pt{\hss\rm Z}}}}}

\def\vev#1{\langle      #1 \rangle}
\def\VEV#1{{\big\langle #1 \big\rangle}}

\def\ME#1#2#3{{\big\langle#1\big\vert#2\big\vert#3\big\rangle}}

     \def\CD{{\cal D}}
     
     \def\CL{{\cal L}}
    \def\CO{{\cal O}} 
    \def\CS{{\cal S}}

\newcommand{\NPBPS}[1]{Nucl.\ Phys.\ (Proc.\ Suppl.) \ {\bf B#1}}
\newcommand{\PL}[1]{Phys.\ Lett.\ {\bf #1}}

\newcommand{\PR}[1]{Phys.\ Rev.\ {\bf #1}}

\begin{document}
\begin{titlepage}
\hskip 10cm \vbox{
\hbox{DFTT 8/2000}
\hbox{February 2000}}
\vskip 0.7cm
\centerline{\Large\bf Lattice Gauge Theories and the} 
\centerline{\Large\bf AdS/CFT Correspondence.}
\vskip 0.4cm
\centerline{\large M. Caselle\footnote{E--mail: {\tt caselle@to.infn.it}}}
\vskip .2cm
\centerline{\small\it Istituto Nazionale di Fisica Nucleare, Sezione di Torino}
\centerline{\small\it  Dipartimento di Fisica
Teorica dell'Universit\`a di Torino}
\centerline{\small\it  via P.Giuria 1, I-10125 Turin,Italy}
\noindent
\vskip 1cm
\begin{abstract}
This is the write-up of a set of lectures 
on the comparison between Lattice Gauge Theories and AdS/CFT results for the
non-perturbative behaviour of non-supersymmetric Yang Mills theories.
These notes are intended for students which are assumed  {\sl not} to be
experts in L.G.T. For this reason 
the first part is devoted to a pedagogical introduction to the Lattice
regularization of QCD. In the second part we discuss some basic features of the
AdS/CFT correspondence
 and compare the results obtained in the non-supersymmetric
limit with those obtained on the Lattice. We discuss in particular the behaviour
of the string tension and of the glueball spectrum.
Lectures delivered at the {\sl School of
Theoretical Physics}~ (S.N.F.T.),~ Parma,~ September 1999. 
\end{abstract}
\end{titlepage}

\setcounter{footnote}{0}
\def\thefootnote{\arabic{footnote}}
\tableofcontents
\vfill\eject
\setcounter{footnote}{0}
\def\thefootnote{\arabic{footnote}}
\section{Introduction}
\label{intro}

Our present understanding of QCD is based on the  
 widely accepted idea that the confining regime of Yang-Mills theories should be
described by some kind of effective string model~\cite{conj,h74}.

This conjecture has by now a very long history. It originates
from two independent observations. 
\begin{itemize}
\item
The first one is of 
phenomenological nature,
and predates the formulation of QCD. 
It is related to the observation that the linearly rising Regge
trajectories in meson spectroscopy can be easily explained assuming a
string-type interaction between the quark and the antiquark. This observation
was at the origin 
of a large amount of papers which
tried to give a consistent quantum description of strings. 

\item
The second one comes from the lattice regularization of pure gauge theories
(LGT in the following) and was realized right after the formulation of QCD.
In the LGT framework one can easily study non-perturbative phenomena, like
those involved in the conjectured string description of $YM$ theories and it is
easy to see that in the strong
coupling limit of pure LGTs the
interquark potential rises linearly, 
and that the chromoelectric flux lines are
confined in a thin ``string like'' flux tube~\cite{wilson}.
\end{itemize} 
Some clear indications were later found that the vacuum expectation value of
Wilson loops could be rewritten as a string functional integral even in the
continuum \cite{polya,gene,nambu}. This led to conjecture  that there exists 
an exact duality between gauge fields and strings \cite{polya}.

However, despite these results, in the following years 
all the attempts to explicitly construct
the conjectured string description of QCD failed.
 In fact it was realized at the beginning of the eighties that 
the strong coupling approximation for the lattice description of
 interquark potential
is plagued by lattice artifacts which make it inadequate for the
continuum theory (this is the famous ``roughening transition'' that we shall
discuss in sect.~\ref{s3.5.1}). 

Since then, while impressive results were
obtained  by means of Montecarlo simulations, very little
progress have been achieved with analytic techniques. Lattice gauge theories
can be solved exactly in two dimensions for any gauge
group, but become unaffordably complex in more than two dimensions, even in
absence of quarks. Moreover, most of the approximation techniques which are
usually successful in dealing with simpler statistical mechanical systems, like
(suitably improved) mean field methods or strong coupling expansions 
turn out to be less useful in the case of LGT.

Several proposals were made during the eighties to 
 overcome these difficulties. In particular, two of them led to rather
interesting results.
\vskip 1.5cm
\begin{description}
\item{\bf Effective string theory}

The first proposal was to assume a milder version of the  
conjecture. This milder version only requires that the
behavior of large Wilson loops is described in the infrared limit by an
effective two-dimensional field theory (hence not a true string theory)
which accounts for the string-like
properties of long chromoelectric flux tubes. 
We shall refer to such a $2d$ field
theory in the following as the ``effective string theory''.
Several interesting results can be obtained in this framework. We shall discuss
them in  detail in sect.~\ref{s3.5} below. Let us anticipate here that the most
interesting feature of this approach is that it leads to predictions which are
in very good agreement with the results of Montecarlo simulations of QCD.
Its major drawback is that it is not consistent at the quantum level.
It is not clear how to extend this effective string description to the
ultraviolet regime, i.e. how to relate it with some kind of
 ``fundamental'' string which is consistent at the quantum level.

\item{\bf Large $N$ limit}

The second proposal was to study the large $N$ limit of $SU(N)$ gauge theories
instead of the phenomenologically relevant $SU(3)$
 model~\cite{h74}. It was shown
that this  large $N$ approximation~\cite{ek} is able to keep the 
whole complexity of the finite $N$ models. 
Unfortunately, even in
the large $N$ limit (despite the fact that some 
major simplifications occur) it is not possible to give exact solution 
(the so called ``Master Field'') to the Lattice SU$(N)$ model and  
 essentially no improvement was made in these last fifteen years also in this
 direction. 

\end{description}

Recently this situation drastically changed thanks to a new, original
 proposal based on the Maldacena conjecture~\cite{m98}
 which relates the large $N$
 expansion of certain supersymmetric gauge theories to the behaviour of
string theory in a  non-trivial geometry. Witten's extension~\cite{w98} of this
conjecture to non-supersymmetric gauge theories, led to the hope of a possible
non-perturbative description also for  large $N$
QCD in four dimensions. In fact in these last months several attempts 
have been made to extract predictions for the string tension and the 
glueball spectrum of large $N$ QCD. 
These predictions have some appealing features, but
also raised serious criticism. 
All the authors agree that some independent test of the applicability of 
the AdS/CFT correspondence to large $N$ QCD is needed. 
This is indeed possible thanks to the impressive progress  of
 montecarlo simulations of LGT which have by now reached stable estimates
 both of  string tension and glueball spectrum for finite $N$ and allow 
 reliable extrapolations to the large $N$ limit.

The aim of these lectures is to
 allow the reader (which is assumed $not$ to be an expert of LGT) to understand
 how these estimates were obtained and to test their reliability. 
We shall also compare
 the ``effective string model'' 
 mentioned above with the string theory which is at the basis of AdS/CFT model. 
The goal is
 to be able not only to accept or reject the AdS/CFT predictions on the basis
 of the LGT results but also, if possible, to
 gain some insight in the AdS/CFT proposal itself. 

To this end we shall devote
 the first part of these lectures (sect.~\ref{s2} and the first part of 
sect.~\ref{s3})
 to an elementary introduction to the lattice
 regularization of QCD, starting form the very beginning. 
Then in the second
 part of the lectures (from sect.~\ref{s3.4} to \ref{s3.7})
 we shall jump to our main object of interest and study in
 some detail the LGT results for the string tension and the glueball spectrum.
 
Unfortunately we shall have to skip several important and interesting hot 
 topics
 of LGT, like the issue of improved (and ``perfect'') actions, that of chiral
 fermions, topological observables, deconfinement transition ....
 We leave the interested reader to the books and review articles listed at the 
 end of the bibliography~\cite{Rebbi}-\cite{latproc} (we tried to make
 the list as complete as possible)
  which summarize the present state of the art in LGT.

The last part of these lectures (sect.s~\ref{s4} and \ref{s5})
 is  devoted to the AdS/CFT correspondence and in
particular to discuss its predictions for the non-perturbative behaviour of
non-supersymmetric Yang-Mills theories. 
Let us stress in this respect that this set of lectures is {\sl not} intended as
 an introduction to the AdS/CFT correspondence
for which we refer to the other lectures delivered at this school and to
two thorough reviews which already exist on
 the subject~\cite{rev_pdv},\cite{rev_agmoo}.
In this lectures we 
shall assume that the reader is already acquainted with the topic 
and shall only remind some basic informations at the beginning of 
sect.~\ref{s4}.

\newpage
\section{Introduction to Lattice Gauge Theories.}
\label{s2}
\subsection{Quantum Field Theories and Statistical Mechanics.}
\label{s2.1}

The modern approach to
 Quantum Field Theories (QFT in the following) is based 
on Feynman's path integral formulation. Using path integrals
 impressive results have been
obtained in the last fifty years in perturbative QFT
 However these methods
require the existence of one (or more) weak coupling parameters in which the
theory can be expanded perturbatively. As such they are not suited for the
analysis of phenomena governed by intrinsically large coupling constants, or 
even worse, with a non-analytic behaviour at the origin, in  the space of
complex coupling parameters. This is exactly the case of QCD, at least as far
as dimensional observables are concerned. To overcome these difficulties a
different regularization was proposed almost thirty years ago by K.
Wilson~\cite{wilson}. Wilson suggested to formulate the theory on a discrete
lattice of points in Euclidean space-time. Such proposal has some very
important advantages:
\begin{itemize}
\item
The path integral becomes a collection of well defined ordinary integrals at
the lattice sites.
\item
The lattice spacing becomes an ultraviolet cut-off. 
\item
As far as the number of sites of the lattice is kept finite all the ultraviolet
divergences are removed and all quantum averages are given by mathematically
well defined expressions, {\sl for any value of the coupling constant}.
\item 
A QFT in $d$ space and 1 time dimensions 
regularized on a lattice becomes equivalent to an equilibrium
statistical mechanics model in $(d+1)$ space dimensions. 
As a consequence one can
study the model with all the tools which are typical of statistical 
mechanics like strong coupling expansion or Montecarlo simulations.
\end{itemize}
Obviously the lattice regularization is not a magic wand and
all the problems which have been overcome  appear again, in some
other form, when we take the continuum limit (we shall deal with this very
delicate issue in sect.~\ref{s2.3}).
 However the main feature of the regularization, i.e. the fact that
it is intrinsically non-perturbative survives in the limit and allows one to
obtain results which could never be obtained with standard perturbative
expansions .

In this section we shall discuss in details the two main steps which allow one
to construct (and extract results from) a lattice regularization of QFT:
\begin{description}
\item{a]}
The translation from Minkowski to Euclidean Quantum Field Theory
\item{b]}
The connection between Euclidean QFT 
 and Statistical Mechanics in the canonical Ensemble.
\end{description}

The starting point for a path integral formulation of QFT is  
 the vacuum to vacuum amplitude (also called the
generating functional) in presence of an external source
$J$
\eq 
Z[J] = \int d \phi e^{
\frac{i}{\hbar}\int dt\int d^3x {\cal L} (\phi)
+ J \phi}~~~~.
\label{zm}
\en
 Correlation functions can be
obtained from this in the standard way by differentiating $\log Z[J]$
with respect to $J$, for example 
\eq
\ME{0}{T[\phi(x) \phi(0)]}{0} = {\delta
\over \delta J(x)} {\delta \over \delta J(0)} \log Z[J] |_{J=0}~~~.
\en

Looking at eq.(\ref{zm}) we immediately recognize the analogy with the
usual expression of the partition function in the canonical ensemble.
The only difference (which is however of great importance) is that in the
exponential of eq.(\ref{zm}) we have an oscillatory term
while the argument in the exponent of a Boltzmann weight
is real. We can bridge this difference with an analytic continuation of 
eq.(\ref{zm}) to imaginary values of the time. This is the well known
 ``Wick rotation''.

\eqa
x_0 &\equiv& t \to -i x_4 \equiv -i \tau \,, \nonumber \\
p_0 &\equiv& E \to i p_4 \,.
\ena
In this way we obtain
\eq 
Z[J] = \int d \phi e^{-\CS_E
+ J \phi}
\label{ze}
\en
 where $\CS_E$ denotes the Euclidean action. We shall discuss in detail its
 form in the Yang Mills case in the next section. At this point we may well
 interpret $\CS_E$ as the Hamiltonian of a static model in four space
 dimensions and
 $Z[J]$ with the corresponding partition function. 
It is far from obvious that we can perform a Wick rotation without problems. On
the continuum this is granted by the good analyticity properties of the
propagator, but on the lattice it imposes some strict constraint on the form of
the discretized action. These constraints are known as reflection positivity
conditions~\cite{os} (and also as  ``Osterwalder and Schrader positivity
conditions'').

The connection between QFT and Statistical mechanics is  a crucial issue
of modern quantum field theory. It has deep, far reaching, consequences in
several physical contexts. Its main implications are summarized in 
tab.~\ref{stat_qft}

\begin{table} 
\begin{center}
\setlength\tabcolsep{0.15cm}
\caption{The equivalence between a Euclidean
field theory and Classical Statistical Mechanics.}
\begin{tabular}{|l|l|}
\hline
Euclidean Field Theory           &  Classical Statistical Mechanics \\
\hline
Vacuum                           & Equilibrium state      \\
Action                           & Hamiltonian                     \\
unit of action $h$               & units of energy $\beta = 1/kT$  \\
Feynman weight for amplitudes    & Boltzmann factor $e^{-\beta H}$ \\
$\qquad e^{-\CS/h}=e^{-\int\CL dt /h}$   &                                 \\
Generating functional       & Partition function $\sum_{conf.} e^{-\beta H}$  \\
$\qquad$ $\int\CD\phi e^{-\CS/h}$  &                                  \\
Vacuum energy                    & Free Energy                      \\
Vacuum expectation value $\ME{0}{\CO}{0}$ &
 Canonical ensemble average $\VEV{\CO}$ \\
Time ordered products            & Ordinary products                \\
Green's functions $\ME{0}{T[\CO_1\ldots\CO_n]}{0}$ &
 Correlation functions $\VEV{\CO_1\ldots\CO_n}$ \\
Mass $M$                         & correlation length $\xi = 1/M$ \\
Mass-gap                         & exponential 
decrease of correlation functions \\
Mass-less excitations            & spin waves  \\
Regularization: cutoff $\Lambda$ & lattice spacing $a$ \\
Renormalization: $\Lambda \to \infty$ & continuum limit $a \to 0$ \\
Changes in the vacuum            & phase transitions \\
\hline
\end{tabular}
\label{stat_qft}
\end{center}
\end{table}
\smallskip
\subsection{Lattice discretization of pure Yang-Mills theories}
\label{s2.2}

The goal of this section is the explicit construction of a lattice
regularization of a gauge theory with gauge group ${\bf G}$.
To this end we need first of all a Wick rotated, Euclidean formulation of the
theory (sect.~\ref{s2.2.1}). Then for its lattice discretization three main
 ingredients are needed: the lattice structure (sect.~\ref{s2.2.2}), the lattice
 definition of the gauge 
variables (sect.~\ref{s2.2.3}) and the action (sect.~\ref{s2.2.4}). 
In each of these steps we have a great amount of
freedom. We shall always choose the simplest option, and leave to exercise 1
the discussion of
possible alternative choices. We shall then check in sect.~\ref{s2.2.5} 
that the proposed action
gives in the ``naive'' continuum limit the expected gauge invariant expression
and add some further remarks on the integration measure (sect.~\ref{s2.2.6}),
 on the fermionic sector (sect.~\ref{s2.2.7}) and on the
 constraints which must be imposed to obtain
 a finite temperature version of the theory (sect.~\ref{s2.2.8}).

\subsubsection{Euclidean Yang-Mills theories}
\label{s2.2.1}

In the following we shall be interested in the lattice formulation of $YM$
theories. The continuum limit of these models is the {\sl Euclidean}
version of $YM$ theories. Let us briefly remind its expression.
The building blocks are the field $A_\mu^i$, $i=1...N$ where $N$ is the number
of the generators of the gauge group. The indices $i,j,...$ run in the space of
the generators of the gauge group.
The Yang Mills action is:
\eq
S_{YM}=\frac14\int d^4x F_{\mu\nu}^iF^{i~\mu\nu}
\en
with 
\eq
F_{\mu\nu}^i=\partial_\mu A_\nu^i -
\partial_\nu A_\mu^i +g f_{ijk}A^j_\mu A^k_\nu
\en
where $g$ is the coupling constant and $f_{ijk}$ are the structure constants of
the gauge group defined by
\eq
[\tau_i,\tau_j]=i f_{ijk}\tau_k~~~.
\en

\subsubsection{The lattice}
\label{s2.2.2}

Let us choose the simplest possible lattice structure: a four dimensional
hypercubic lattice ${\bf \Lambda}$ of size $L$ in the four directions. 
Let us denote
the sites of the lattice with $n\equiv (n_0,n_1,n_2,n_3)$ and with $\hat\mu$
the unit vector in the $\mu$ direction $(\mu=0,1,2,3)$. 
We shall often call
in the following the $0$ direction as time-like and the other three as
space-like, but let us stress that, due to the Wick rotation, all four
directions are exactly on the same ground (we shall make use of this symmetry
in the following). ${\bf \Lambda}$ contains $N_s~=~L^4$ sites,
 $N_l~=~4~L^4$ links and $N_p~=~6~L^4$ plaquettes. We shall denote with $n_\mu$
 the link starting form $n$ and pointing in the positive $\mu$ direction i.e
 the link joining the two sites $n$ and $n+\hat\mu$. Similarly
 $n_{\mu,\nu}$ denotes the plaquette joining the four sites $n$, $n+\hat\mu$,
 $n+\hat\mu+\hat\nu$, $n+\hat\nu$. 
We shall denote with $a$ in the following the ``lattice
 spacing'' i.e. the separation between two nearby sites. 
\subsubsection{The gauge field.}
\label{s2.2.3}
There is a  standard recipe to choose the lattice analog of the {\sl
bosonic} fields of a continuum QFT. One must define the scalars on the sites, 
the vectors on the links and the two index tensors on the plaquettes of the 
lattice. This recipe can be simply understood if we rewrite the QFT in terms of
differential forms. The discrete analog of a p-form is a p-simplex. In whole
generality this rule can be stated as follows:

``The lattice analog of a tensor field of degree $k$ is a  function which
takes values on the $k$ simplexes of the lattice''

Thus the vector potential $A_\mu(x)$ must be defined on the links of the 
lattice.
The simplest choice, which ensures both gauge invariance and a smooth continuum
limit is to put in each link $n_\mu$ an  element of the 
gauge group:  $U_\mu(n)\in{\bf G}$ such that 
\eq
U_\mu(n)=e^{cA_\mu(n)}
\en
where $c$ is a suitable constant which we shall discuss below. 

A nice, intuitive, way to understand this choice is the following:
Imagine that in each site $n$ does exist an internal space $E$ (on which the
gauge group $\bf G$ acts in a non trivial way). Let us assume that the
reference frame $E_n$ on $E$ changes from site to site. Then we can interpret
$U_\mu(n)$ as the transformation which relates the two 
nearby reference frames $E_n$ and $E_{n+\hat\mu}$. An immediate
 consequence of this
picture is that we must impose, for consistency:
\eq
  U_{-\mu}(n+\hat\mu)=U_\mu^{-1}(n)
\en
In this framework a gauge transform is simply an arbitrary rotation, site by
site of the reference frames $E_n$. Let us denote these transformations
 as $V(n)\in{\bf G}$. It is clear that the effect on $U_\mu(n)$ of such
 transformations is the following
\eq
U_\mu(n)~~ \to~~ V(n)~U_\mu(n)~V^{-1}(n+\hat\mu)
\label{gaugetransf}
\en
Thus, as expected, the single variable $U_\mu(n)$ is {\sl not} gauge invariant.
The simplest way 
 to construct, out of the gauge variables $U_\mu(n)$, gauge
invariant observables, is to choose a closed path $\gamma$ on the lattice
and then construct 
\eq
W(\gamma)={\rm Tr} \prod_{n_\mu\in\gamma}U_\mu(n)
\en
(where the product is assumed to be ordered along the path $\gamma$).
This observable is usually called ``Wilson loop''.

\subsubsection{The action}
\label{s2.2.4}
Obviously, the main requirement that we must impose on the discretized version
of the action is that it must be gauge invariant. Among all the possible Wilson
loops the simplest one is the product of the four link variables around a
plaquette. If we sum these elementary Wilson loops over all the plaquettes of
the lattice we obtain an expression which is invariant with respect to the
discrete subgroups of the translational and rotational symmetries which 
survive in the lattice discretization. 
This was the original Wilson proposal for the lattice
discretization of the gauge invariant action. 
Such an expression defines a perfectly consistent gauge invariant model for any
group $\bf G$ on the lattice and is a good candidate to define a translational
and rotational invariant gauge theory also in the continuum limit. 
At this stage we have no constraint on  $\bf G$
which can well be a finite, discrete group. For instance several interesting
results have been obtained in the case in which  ${\bf G}=Z_2$ (the ``gauge
Ising model''). However since we are interested in constructing the lattice 
version of
Yang-Mills theories we must concentrate on the case in which $\bf G$ is
continuous Lie group. Even if the physically interesting case is
 ${\bf G}=SU(3)$, we shall study  in  the following the general 
case ${\bf G} = SU(N)$ with $N\geq2$. This extension essentially does not add
any further complication and allows to study the limit $N\to\infty$ which is
essential if we aim to compare our lattice results with the AdS/CFT
predictions.

\begin{figure}[htb]
\centerline{\epsfxsize=5truecm\epsffile{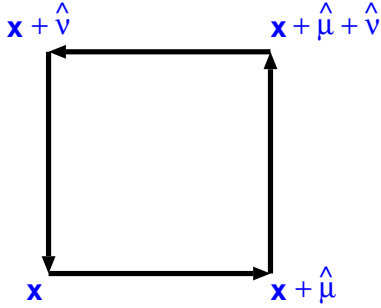}}
\caption{The Wilson action.}
\label{ads_plaq}
\end{figure}

Let us see it in detail. Let us define with $U_{\mu\nu}(n)$
 the product of the four link variables around the plaquette $n_{\mu\nu}$ (see
 fig.~\ref{ads_plaq})

\eq
U_{\mu\nu}(n)=\left\{U_\mu(n)
U_\nu(n+\hat\mu)U_{-\mu}(n+\hat\mu+\hat\nu)U_{-\nu}(n+\hat\nu)\right\}
\label{umunu}
\en
The Wilson action is
\eq
S_W=-\frac{\beta}{2N}\sum_{n,\mu\nu} {\rm Re~ Tr}\left\{U_{\mu\nu}(n)
\right\}~~~~,
\label{w1}
\en
where we have introduced the $\frac{\beta}{2N}$ constant 
 in front of the action  for future convenience. Notice that since the sum over
 $\mu$ and $\nu$ is unrestricted  all the plaquettes of the lattice are 
counted twice. We shall check in the next section
 that this proposal indeed gives in the continuum
limit the correct gauge invariant action.

\begin{itemize}
\item
{\bf Exercise 1: discuss some possible generalizations of the lattice
discretization of $SU(N)$ $YM$ theories.}
In the above derivation we chose at each step the simplest possible option.
However there are infinitely many different lattice regularizations which lead
to the same continuum limit of eq.(\ref{contYM}). Discuss some of these possible
generalizations. 
\end{itemize}

\subsubsection{``Naive'' continuum limit.}
\label{s2.2.5}
Let us call ${\cal G}$ the Lie algebra associated with $\bf G$. Let us assume
that ${\cal G}$ has  $N$ generators $\tau_1 \cdots \tau_N$. Then we can define:
\eq
U_\mu(n)\equiv e^{iB_\mu(n)}
\en
with
\eq
B_\mu(n)\equiv a g \tau_i A_\mu^i(n)
\en
$a$ being the lattice spacing and  $g$ a suitable coupling constant.
As we shall see 
the $A_\mu^i$ functions will become in the continuum limit the standard 
vector potential fields of the Yang Mills theory.

We may expand the fields appearing in the Wilson action (keeping only the first
order in $a$) as follows.
\eqa
B_\nu(n+\hat\mu) &\sim& B_\nu(n) + a \nabla_\mu B_\nu(n) \nonumber \\
B_{-\mu}(n+\hat\mu+\hat\nu) &\equiv& 
-B_\mu(n+\hat\nu)~~\sim~~ -B_\mu(n) - a \nabla_\nu
 B_\mu(n) \nonumber \\
B_{-\nu}(n+\hat\nu) &\equiv& - B_\nu(n)
\ena
 where we have denoted with
\eq
\nabla_\nu f(n)\equiv \frac{ f(n+\hat\nu) -f(n)}{a}
\en
the finite difference on the lattice, whose continuum limit is the partial
derivative:
\eq
\nabla_\nu f(n) ~~\to~~ \partial_\nu f(x)~~~~~.
\en
From the above expansions we obtain:
\eq
U_{\mu\nu}(n)~~\sim~~
e^{iB_\mu(n)}
e^{i(B_\nu(n)+a\nabla_\mu B_\nu(n))}
e^{-i(B_\mu(n)+a\nabla_\nu B_\mu(n))}
e^{-iB_\nu(n)}
\en
Let us use at this point the Baker-Hausdorff formula, which at the first order
is:
\eq
e^x e^y= e^{x+y +\frac12[x,y]}
\en
Keeping in the expansion only terms up to $O(a^2)$ (remember that $B_\mu$
is of order $a$) we find
\eq
U_{\mu\nu}~~\sim~~
e^{\left\{ia(\nabla_\mu B_\nu-\nabla_\nu B_\mu)
-[B_\mu,B\nu]\right\}}\equiv 
e^{ia^2gF_{\mu\nu}}
\en
where we have neglected for simplicity the argument $n$ and we have defined:
\eq
F_{\mu\nu}\equiv \nabla_\mu A_\nu - \nabla_\nu A_\mu + ig [A_\mu,A_\nu]
\en
and $A_\mu\equiv \tau_iA^i_\mu$.

Let us insert this result in eq.(\ref{w1}), and expand in powers of $a$.
we find:
\eq
S_W\sim -\frac{\beta}{2N}\sum_{n,\mu\nu} 
{\rm Re~~ Tr} \left({\bf 1} + ia^2gF_{\mu\nu} - \frac12 a^4g^2F^2_{\mu\nu} 
\right)
\en
We can always parameterize the $SU(N)$  generators 
 so as to have
\eq 
{\rm Tr} (\tau_i)=0,~~~~~~~~~~
 {\rm Tr} (\tau_i\tau_j)=\frac12 \delta_{ij}~~~~.
\en
while ${\rm Tr}~ {\bf 1}$ only gives an irrelevant constant which can be
neglected. In this way we obtain
\eq
S_W=
\frac{\beta a^4 g^2}{8N}\sum_{n,\mu\nu} F^i_{\mu\nu}F^{i~\mu\nu} + O(a^5)
\label{contYM}
\en
In the naive limit $a\to 0$ we can set $\sum_{n}\to \int\frac{d^4x}{a^4}$
(for a four-dimensional lattice) and we see that the dominant term in the $a\to
0$ limit becomes 
the standard expression of the pure $YM$ action if we set
\eq
\beta=\frac{2N}{g^2}
\en
From this position we see that $\beta$ is proportional to the inverse of $g^2$.
It is also easy to see, from dimensional arguments that in four dimensions the
coupling constant $g$ is adimensional while for 3d $YM$ theories 
$g^2$ has the dimensions of a mass.

\subsubsection{The partition function.}
\label{s2.2.6}
As mentioned above, our real interest is not in the gauge action but in the
partition function (or generating functional in the QFT language) $Z$.
In constructing $Z$ we must address the problem of the integration measure
(and the related problem of gauge fixing this integration).
Here we see one of the most interesting advantages of  the lattice
regularization. The integrals involved in the construction of $Z$ are site by
site (or better, in the case of a gauge theory, link by link) ordinary
integrals. We have a natural choice for the integration measure: the Haar
measure (i.e. the invariant measure over the group manifold $dU_\mu(n)$).
We have:
\eq
Z=\int \prod_{n,\mu} dU_\mu(n) e^{-S_W}
\en
and similarly, for a generic expectation value we have
\eq
\vev{{\cal O}}=\frac{\int \prod_{n,\mu} dU_\mu(n) {\cal O}(U_\mu(n))e^{-S_W}}{Z}
\label{vev}
\en

A remarkable consequence of these definitions is that, since the integration is
made, link by link, over the whole group manifold {\sl all} the gauge
equivalent configurations (see eq.(\ref{gaugetransf})) are automatically included in the sum with the
correct weight. {\sl Contrary to the continuum case, on the lattice, the
integration over the pure gauge degrees of freedom does not make quantum
averages ill defined}. In the lattice regularization there is no need to fix
the gauge: all the quantum averages are by construction gauge invariant.

\subsubsection{Fermions.}
\label{s2.2.7}
In the previous sections we studied the lattice discretization of 
pure $YM$ theories. In full QCD we should also take into account the quarks.
However putting fermions on the lattice is a rather non trivial issue.
Moreover, once a consistent discretization of the theory is obtained if we try
to integrate out the fermions from the Lagrangian we obtain a determinant of the
gauge fields  which is very difficult to handle in
Montecarlo simulations. In view of this consideration it has become a common
habit to organize lattice gauge theories in three levels of increasing
complexity. 
\begin{itemize}
\item
In the first level we find pure $YM$ theories. These models are by
now rather well understood and precise Montecarlo results exist for the
continuum limit of several quantities, among these the most interesting ones
are the glueballs since we may rather safely expect that their mass should not
be too affected by the absence of quarks. 
\item
 The second level is the so called
``quenched QCD'' in which the quarks are explicitly added into the game, thus
giving the possibility to explore several new observables (for instance
 the meson spectrum) which are of great interest for the phenomenology, but
 they are kept quenched, i.e. the determinant mentioned above is simply
 neglected. This means that in the partition function the quarks are treated as
 classical quantities or equivalently that we are neglecting quark loops in our
 calculations. Also for quenched QCD stable and reliable result have been
 obtained from Montecarlo simulations. However there are effects, like the
 string breaking at large distance, which are a typical consequence of quarks
 loops and that cannot be observed in the quenched approximation. Moreover it
 is by now clear that several quantities of great physical interest (like
 the meson spectrum or the temperature of the chiral phase transition) are
 heavily affected by such approximation.
\item
 The third level is that in which full
 QCD, with dynamical fermions is discretized on the lattice. In this last case
 Montecarlo simulations are still at a preliminary stage and a few years (and
 the next generation of supercomputers) will be needed before we may reach stable
 and reliable results also in this case.
\end{itemize}
 Luckily enough,
 the predictions of the AdS/CFT correspondence,
 whose  comparison with the lattice
 results is the main goal of these lectures, refer to the pure $YM$ theories.
 Thus on the lattice side of the comparison we may rely on very stable and 
 trustable numbers. Moreover in pure $YM$ theories extensive simulations exist
in three dimensions also for values of $N$ larger than 3
 and  reliable large $N$
limits for various quantities of interest can be obtained~\cite{teper1}. 
In four dimensions the results of these large $N$ limits are still
preliminary~\cite{teper2},
but nevertheless they are stable enough to allow for the comparison with the
AdS/CFT results.
In the following, both in the lattice sections and in the
 AdS/CFT ones we shall make an effort to keep well distinct those results which
 refer to pure Yang Mills theories from those which refer to full QCD.

\subsubsection{Finite temperature LGT.}
\label{s2.2.8}
We shall see in sect.~\ref{s4} that in the framework of the AdS/CFT
correspondence a
very important role is played by the ``temperature'' of the theory. It turns
out that it is only at high temperature that we can get rid of the
supersymmetry and obtain non-supersymmetric  Yang-Mills like theories. It is
thus important to understand how can we describe finite temperature on the
lattice.

The model that we have defined and studied in the previous sections describes
$YM$ theories at strictly zero physical temperature. The parameter which
 in the statistical
mechanics counterpart of the model plays the role of the temperature 
becomes in LGT the coupling constant of theory. The discussion of 
sect.~\ref{s2.1} can be easily extended so as to implement 
a non-zero temperature also in LGT. The main ingredient is that periodic
boundary conditions must be imposed in the ``time direction''. Then it can be
shown that the inverse of the lattice size in this direction is proportional to
the temperature of the theory. 

In general, for technical reason, one imposes
periodic boundary conditions in all the $d$ directions, but one also takes care
to choose the lattice length much larger that the typical correlation length of
the theory, so as to make  negligible the effect of the boundary conditions. On
the contrary if we want to see the effects of the finite temperature in the
theory, the lattice size in the ``time'' direction must be chosen of the 
same order of the correlation length. If we increase the lattice size in the
``time'' direction then we smoothly reach the zero temperature limit, which is
effectively reached when the effects of the periodic boundary condition become
negligible. Thus typically a finite temperature discretization requires
asymmetric lattices, with one direction  much shorter than the others.
As a consequence the original equivalence of space
and time directions which is a typical feature of
 Euclidean QFT is lost. 
In finite temperature LGT (FTLGT in the following), 
we have a very precise notion
of ``time'' which is the compactified direction 
proportional to the inverse of the temperature. 

Due to the presence of periodic boundary conditions a new class of observables
exists in FTLGT i.e. the loops which close winding around the 
 compactified time direction. These are usually called Polyakov loops. 
The discussion of FTLGT (apart from a few issues that we shall address in 
sect.~\ref{s3}) is beyond the scope of these lectures. We refer to the
reviews in the bibliography for further details.

\subsection{Continuum limit}
\label{s2.3}
It is clear that by simply sending $a\to 0$ as in the previous section we do
not obtain a meaningful continuum limit. In particular, all the dimensional
quantities, which will be proportional to a non-zero power of $a$ will  go to
zero or infinity. This is the meaning of the word ``naive'' used above.

Let us study this problem in more detail.
Let us take a physical observable ${\cal O}$ of dimensions $d_{\cal O}$ in
units of the lattice spacing. Let us assume that we have in some way calculated
the mean value of ${\cal O}$ in the lattice regularized version of the theory.
The result of this calculation will take the form:
\eq
\vev{{\cal O}} = a^{d_{\cal O}} f_{\cal O}(g)
\en
where $f_{\cal O}(g)$ is a suitable function of the coupling constant of the
theory (in general of all the parameters of the model if they are more than
one). For instance ${\cal O}$ could be one of the
 correlation lengths of the model (i.e. the
inverse of the mass of one of the states of the theory). In this case 
$d_{\cal O}=1$ and $f_{\cal O}(g)$ measures the correlation length {\sl in
units of the lattice spacing}, for the particular value $g$ of the coupling
constant. It is now clear that if we simply send $a\to 0$ we obtain 
 the trivial result $\vev{{\cal O}} = 0$. In order to have a meaningful
 continuum limit we must change $g$ at the same time as $a$ is set to zero in
 such a way as to make the observable to approach a well defined finite value
 in the limit. In the example this means that we must tune $g$ to a critical
 value $g_c$ in which the correlation length measured  in
units of the lattice spacing goes to infinity.

 From a statistical mechanics
point of view this implies that at $g=g_c$ the system must undergoes a
continuous phase transition. This is a mandatory requirement for a non-trivial
continuum limit. From the point of view of QFT we have a nice interpretation of
this constraint. While the lattice discretization gives a way to {\sl
regularize} the theory. The process of tuning $g$ to its critical value, thus
removing the cut-off, corresponds to the {\sl renormalization} of the theory.

This process is highly non trivial, since for all the physical quantities
 ${\cal O}_i$
that we may define in the model we must require a well defined continuum limit.
If 
$f_i(g)$ is the function which measures the value of
 ${\cal O}_i$  in units of the  lattice spacing, then we must require that 
as $g\to g_c$ {\sl all} the functions $f_i(g)$ go to zero or infinity
(depending on the sign of $d_{{\cal O}_i}$) {\sl in such a way that the same
rate of approach of $g$ to $g_c$ which makes the correlation length $\xi$ tend
to a constant value also makes {\bf all} the observables  ${\cal O}_i$
tend to constant values}. This stringent requirement is better understood in
the framework of the Renormalization Group approach and is
 commonly summarized, by
saying that the critical point $g_c$ must be a {\sl scaling} critical point.

It is clear from the above discussion that a meaningful continuum limit
requires a precise functional relationship between $a$ and $g$. However in
principle we can even ignore such a dependence. The notion of scaling defined
above allows to reach a well defined continuum limit even if we do not know how
to fix $g$ as a function of $a$. If we know 
the physical value of one of the observables of the theory, say the mass $m_e$
of a particle ``e'' which is easily accessible from the experiments, then we may
set the overall scale of the theory as follows
\eq
m_e=\frac{1}{a(g)\xi_e(g)}
\en
where $\xi_e(g)$ is the particular correlation length related to the particle
``e''.
Then  the continuum limit value of any other dimensional quantity  of the
 theory, like for instance the masses $m_i$ of other particles, can be obtained
 as
\eq
m_i=\lim_{a\to 0}\frac{1}{a(g)\xi_i(g)}~=~\lim_{g\to g_c}
\frac{\xi_e(g)}{\xi_i(g)}m_e
\en
This is the power of scaling!

In some cases it may happen that, on top of the above relations, we also
have some independent way to fix asymptotically (i.e. in the vicinity of the
critical point) the relationship between $g$ and $a$.
 This is the exactly  
 the case for non-abelian gauge theories, for which asymptotic
freedom tells us that $g_c=0$ and perturbative methods can be used. This leads
to the following well known expression, for a pure $SU(N)$ gauge theory in four
dimensions:
\eq
a=\frac{1}{\Lambda}f(g)
\label{cs0}
\en
with
\eq
f(g)=(g^2\beta_0)^{\beta_1/(2\beta_0)}e^{-1/(2\beta_0 g^2)} (1+ O(g^2))
\label{cs}
\en
where 
\eq
\beta_0=\frac {11 N}{48\pi^2},~~~
\beta_1=\frac {34}{3}
\left[\frac{N}{16\pi^2}\right]^2
\en
are the first two coefficients of the Callan-Symanzik function $\beta(g)$ and
$\Lambda$ is a scale parameter, which does not have a direct physical meaning
and in general depends on the renormalization scheme that we have chosen (for
instance it may depend on the type of lattice that we have chosen).

One usually refers to this relation (and to the procedure of taking the
continuum limit following  eq.(\ref{cs}))
 as ``asymptotic scaling'' to stress the
fact that one is using more informations than those simply implied by the 
scaling property.

Since it will play an important role in the following it is worthwhile to
discuss in detail how one should
 use eq.(\ref{cs}). To this end let us continue with the above
example, and let us assume again that  ${\cal O}$ is the correlation length of
the theory (i.e the inverse of the lowest glueball). Let us measure it on the
lattice for various values of $g$ in the scaling region in units of the lattice
spacing $a$ and let us call these numbers (which at this point are pure
adimensional numbers) $\xi(g)$. We have
\eq
{\cal O}=a\xi(g)
\en
Let us now insert eq.(\ref{cs0}), we find
\eq
{\cal O}=\frac{f(g)\xi(g)}{\Lambda}
\en
Our goal is to find a finite continuum limit value (let us call it 
$\frac{\xi_0}{\Lambda}$) for
${\cal O}$. This implies that $\xi(g)$ must scale as:
\eq
\xi(g)=\frac{\xi_0}{f(g)}
\label{cs21}
\en
If this condition is fulfilled by the data then we may say that the lowest
glueball has a mass in the continuum limit whose values is
\eq
m=\frac1\xi_0\Lambda
\label{cs22}
\en
if we are able to fix the value of $\Lambda$ in $MeV$ (and this can be done,
for instance, by comparing the string tension
evaluated on the lattice with
the physical value of the string tension obtained from the
spectroscopy of the heavy quarkonia) then eq.(\ref{cs22}) will give us the
value in $MeV$ of the lowest glueball. The precision of this prediction will
only be limited by the uncertainty in the determination of $\Lambda$ in $MeV$
and by the statistical and systematic errors which affect the estimate of the
amplitude $\xi_0$ from a fit to the data $\xi(g)$ according to eq.(\ref{cs21}).

Let us conclude this section by noticing that
the functional dependence on $g$
 of eq.s(\ref{cs0}),(\ref{cs}) is a direct consequence of the fact
that the coupling $g^2$ in this case is adimensional. In fact the 
scaling behaviour for $SU(N)$ gauge theories in three dimensions is completely 
different. In this case it is $g^2$ itself (which has the dimensions of a 
mass) which sets the overall scale for the theory. This means that near the
continuum limit a physical observable with the dimensions of a mass like 
for instance the inverse of the correlation length
 that we studied above, can be written as a series in
powers of $g^2a$.
\eq
\frac{1}{a\xi(g)}=mg^2(1+m_1g^2a+m_2g^4a^2+\cdots)
\en
where $m$ is the continuum limit value of the mass and the constants
 $m_1,m_2\cdots$ measure the finite $a$ corrections to the scaling behaviour.

\newpage

\section{Extracting physics from the lattice.}
\label{s3}

In order to extract some physical results from the lattice regularized model 
we need three steps. 
\begin{itemize}
\item First, we
must  define the lattice version of the quantities in which
we are interested. In general there is not a unique prescription to do this. As
in the previous section we shall always choose the simplest lattice realization
and shall then comment on the possible generalizations.
\item Second, 
we must use some  non perturbative technique to
extract the expectation value of these operators for a (possibly wide) set of
values of the parameters of the model (in the simplest case only the coupling
constant $g$). In some very exceptional situation (like 2d $YM$ theories) exact
results can be obtained with analytic techniques, but in general some
approximate method is needed. The  most popular tool is the Montecarlo
simulation, however in some situations also strong coupling expansions can give
reliable results.
\item Third,
 one must test that the $g$-dependence of the measured quantities
scales according to eq.(\ref{cs}). If this condition is fulfilled then one can
set $a\to 0$ and extract in the continuum limit the value of the observable
under study, in units of the physical scale $\Lambda$. 
\end{itemize}
Let us see these steps in detail
\subsection{Lattice observables.}
\label{s3.1}
There are several quantities which are accessible on the lattice. For the
reason discussed in the introduction we shall concentrate in the following
only on two of them: the string tension and the glueball spectrum. 
\subsubsection{The string tension $\sigma$}
\label{s3.1.1}
Phenomenologically we know that the quark and the antiquark in a meson are tied
together by a linearly rising potential. The simplest way to describe such a
behaviour is to assume that the infrared regime of QCD is described by an
``effective'' string (which, as we shall see, is very different from the one
which lives in AdS) which joins together quark and antiquark. This is the
origin of the name``string tension'' to describe the strength of the rising
potential (we shall discuss
in detail this effective string picture in sect.~\ref{s3.5}).

 In the real world the
best set up to extract experimental informations on the string tension $\sigma$
 is represented by the spectrum of the heavy quarkonia where,
 thanks to the large 
masses of the quarks, the quark-antiquark pair can be studied with 
non-relativistic techniques. Suitable potential models can be used to fit the
spectrum and in this way an  experimental estimate for $\sigma$ can
be extracted. 

\begin{figure}[htb]
\centerline{\epsfxsize=8truecm\epsffile{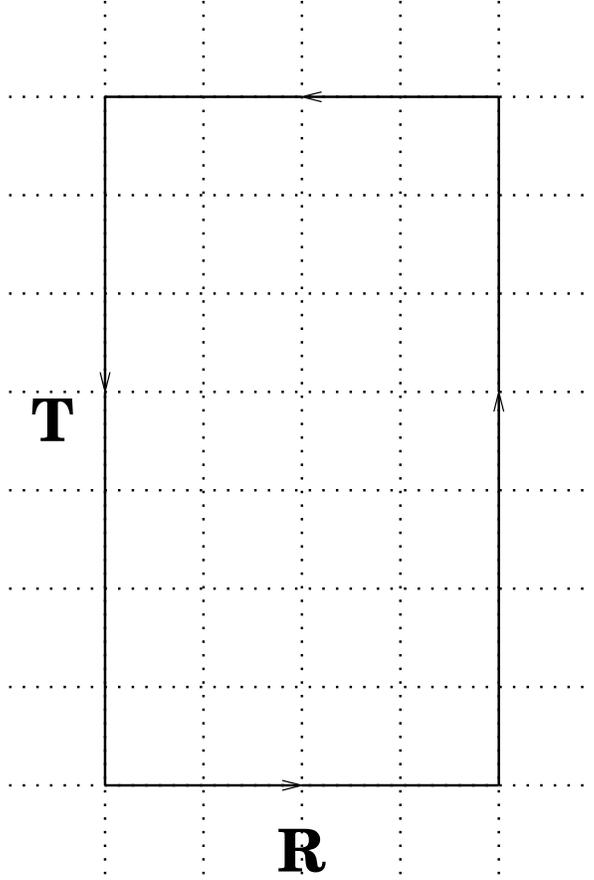}}
\caption{The Wilson loop.}
\label{ads_wilson}
\end{figure}

On the lattice the simplest way to mimic the quark-antiquark pair is to study
the mean value of a
large rectangular Wilson loop (say of sizes $R\times T$). Let us denote it as 
$\vev{W(R,T)}$ (see fig.~\ref{ads_wilson}). Let us also assume that
T is a segment in the ``time'' direction (remember however that the notion of
time direction is purely conventional in our Euclidean framework) from the time
$t_0$ to the time $t_0+T$, and $R$ a segment in any one of the space
directions.

The physical
interpretation of $\vev{W(R,T)}$ 
is that it represents the variation of the free energy due to the
creation at the time $t_0$ of a 
quark and an antiquark  which are instantaneously
moved at a distance $R$ from each other, keep their position
for a time T and finally annihilate at the instant $t_0+T$. According to this
description we expect  for large~$T$  
\eq
\vev{W(R,T)}\sim e^{-T~ V(R)}
\en
where $V(R)$ is the potential energy of the quark-antiquark pair.
The interquark potential is thus given by:
\be
V(R)=-\lim_{T\to\infty} \frac{1}{T}\log\Br W(R,T)\kt
\label{lu0}
\ee

If we assume that $V(R)$ is dominated by the linear term $\sigma R$ then we end
up with the celebrated ``area law'' for the Wilson loop:
\eq
<W(R,T)>=e^{-\sigma RT+p(R+T)+k}~.
\label{area}
\en
 The area term is responsible for confinement 
while the perimeter and constant terms are non universal contributions related
to the discretization procedure. When one takes the $T\to\infty$ limit to
extract the potential (see eq.(\ref{lu0})) the constant term disappears and the
perimeter one gives a normalization constant $V_0$ which can be easily
 fixed from a fit. The physically important quantity is the coefficient of the
 area term which represents the lattice estimate 
  of the string tension which we were looking for.

We shall see below  that the real  expression for 
$\vev{W(R,T)}$ is slightly more complicated, but in any case the dominant term
is the string tension, which is given by
\eq
\sigma=-\lim_{T\to\infty}\frac{1}{RT} log(\vev{W(R,T)})
\label{sigdef}
\en
It is easy to see~\footnote{All the 
dimensional quantities discussed in this section 
$R,T, \sigma$ and $V(R)$ are measured in units of $a$, but here and in the
following  we shall omit the factors of $a$ (which can be easily deduced from a
dimensional analysis) to avoid a too
heavy notation.} that $\sigma$ has dimensions of $a^{-2}$
 and as a  consequence its expected scaling behaviour in $d=4$
is, according to the discussion at the end of sect.~\ref{s2.3},
\eq
\sigma(g)=\sigma_0 f^2(g)
\en
where $f(g)$ is given by eq.(\ref{cs}). From the numerical value of
 $\sigma_0$ we may then obtain the continuum limit value $\sigma_0\Lambda^2$
 (in units of $MeV^2$ once we have fixed the value of $\Lambda$) for
 the string tension.

\subsubsection{Glueballs.}
\label{s3.1.2}
As it is well known, pure gauge theories 
 have a rich spectrum of massive states which are called glueballs. 
This is one of the most remarkable effects of quantization since
classical gauge field 
theories do not contain any mass terms and are scale-invariant. Moreover it is a
truly non-perturbative effect: in perturbation theory the gluon propagator
remains massless to all orders. 

The lattice offers a perfect 
framework to study the
glueball spectrum and in fact on the lattice massive states of pure gauge
models arise in a very natural way. One must select two elementary 
"space-like" plaquettes at two different positions in the ``time'' direction.
Looking at the large distance  (i.e. large time) behaviour
(with a Montecarlo simulation or with a strong coupling expansion)
of the connected correlator of the two plaquettes one immediately
recognizes the exponential decay which denotes the presence of a massive state.
From this behaviour one can extract the correlation length, whose inverse is
the mass of the lowest glueball (the $0^{++}$ state).
\eq
\vev{{\rm Tr}(U_{ij}({\bf x},t_1))~{\rm Tr}(U_{ij}({\bf x},t_2))} \sim
e^{-M|t_1-t_2|}
\en
where $(U_{ij}({\bf x},t_1))$ is the plaquette (with spacelike indices $(i,j)$)
located in the point $({\bf x},t_1)$ of the lattice, and $M$ is the mass of the
lowest glueball.

 It is nice to observe
that the exponential decay is already visible (without the need of a MC
simulation) in the very first order of a
 strong coupling expansion, a very simple calculation that we shall perform
 below as an exercise on strong coupling expansion. However for the non
 abelian gauge theories which are of physical relevance, the SC series are
 limited to rather few terms and allow to obtain only a rough estimate of the
spectrum. If one is interested in comparing the lattice with possible
experimental candidates it is mandatory to use MC simulations.

In order to extract good estimates for higher states of the
 spectrum one must study connected correlators (in the time direction)
of more complicated combinations
 of space-like Wilson loops of suitable shapes. 
These combinations are chosen so as to match 
the symmetry properties of the glueball (which is encoded in
 the notation $J^{PC}$). This is a rather subtle issue since on the
 lattice the rotation group is broken to its cubic subgroup. This has two
 relevant consequences:
\begin{description}
\item{1]}
Some of the irreducible representations
 of the rotation group are not any more irreducible with respect to the cubic
 subgroup and hence split in several components which on the lattice, in
 principle, correspond to different massive states. These masses must coincide
 as $a \to 0$ if we want to recover the correct continuum theory.
This represents a non trivial consistency test  of the continuum limit
extrapolation. For finite values of $a$ the splitting between these 
``fragments''
 of the same state gives an estimate of the relevance of lattice artifacts.
\item{2]}
Conversely, since there is only a finite number of irreducible representations
in the cubic group, any one of them must contain infinitely many irreps of the
rotation group. In particular all values of $J$ (mod 4) coincide in the cubic
subgroup (they can be recognized as metastable states following the suggestions
of ref~\cite{tj}).
\end{description}

 Constructing the correct identifications between  the 
continuum states and their lattice realizations turns out to be
a non-trivial (and very instructive) exercise of group theory. We shall discuss
as an example in exercise 2 the construction in the case of
 the $SU(2)$ theory in (2+1) dimensions. Even if this is the simplest
possible situation it is 
 already complex enough to show all the subtleties of the problem.
  The generalization to other values of $N$ and to the 
(3+1) dimensional case can be found, for instance, in~\cite{MM}.
 It is also possible to disentangle
glueballs with the same $J^{PC}$ but different radial quantum numbers. This is
very important since in general the first state above the $0^{++}$ is
its first radial excitation and not a glueball of different angular momentum.
We can  summarize the discussion of this section (and the analysis performed in
exercise~2) as follows
\begin{itemize}

\item
For any glueball state of quantum numbers $J^{PC}$ with $J<4$
 it does exist a lattice representation in terms of suitable combinations of
 spacelike Wilson loops of various shapes. These combinations can be constructed
 by using group theory.  Let us call them $W(J^{PC})$
\item
 each $W(J^{PC})$ gives a lattice realization for a whole family
of glueball states with angular momentum $J$ (mod 4). Looking at the
 large distance (in the ``time'' direction) behaviour of the connected 
correlator of $W(J^{PC})$ we may extract the one of lowest mass. Higher
glueballs can be seen as exponentially suppressed
corrections in the correlator or as  metastable states. 
\item
This representation is not unique, in general there are infinitely many
combinations of spacelike Wilson loops with the same symmetry properties.
By using some variational method we can select those combinations which enhance
the particular higher mass state in which we are interested (say the first radial
excitation of the $0^{++}$) and thus measure its mass.

\end{itemize}
\vskip1cm
\begin{itemize}
\item {\bf Exercise 2: group theoretical analysis of the 
glueball states for the $SU(2)$ LGT in $d=3$.}
\end{itemize}

\subsection{Strong Coupling Expansions}
\label{s3.2}
In  sect.~\ref{s2.1} we have shown that there is a correspondence between
QFT  and Statistical Mechanics. 
In particular we can interpret the lattice
regularized $YM$ theories as a peculiar statistical model in which $g^2$ plays
the role of the temperature. One of the most powerful tools to
 study statistical models are the high temperature expansions (i.e.
 perturbative expansions in the inverse of the temperature). The main
 ingredient in this game is the expansion of the Boltzmann factor of the model
 on the character basis. In such a basis it becomes very simple to
 perform, order by order, the sum over all the possible configurations of the
 model which appear in the partition function and in the correlators.
Moreover, by using the orthogonality properties of
 the characters, a set of rules can be constructed which greatly simplify the
 terms in the expansion. 
The final result can be written as
 a series in powers of $\beta$ i.e. perturbative in $1/T$ as desired.

 It is easy to export this technique to LGT. The important point is that, thanks
to the identification between temperature in statistical mechanics
 and coupling constant $g^2$ in LGT, the high $T$
expansion becomes in LGT a strong coupling expansion, i.e. an expansion in the
inverse of the coupling constant. But this is exactly what we need to study the
non-perturbative physics of $YM$ theories, and in fact in the strong coupling
limit all the features that we expect  to find in the theory (and have 
never been able to proof), like the linear confinement and a nonzero
mass gap for the glueballs can be explicitly shown. 

In principle,
if we could push the strong coupling
expansion (which is centered in $g^2=\infty$) up to the scaling region (small
values of $g^2$) we would reach the long sought
 continuum limit  description of the
non-perturbative physics of $YM$ theories. Unfortunately this seems a too
difficult task. In some cases, like for the Wilson loop that we shall discuss
in detail below, it can be shown that the task is actually impossible and
that the scaling region is separated form the strong coupling region by a phase
transition, the roughening transition, which cannot be overcome. In some other
cases (like for the glueball masses) the problem is that
too many orders in the strong coupling expansion are needed to reach the
scaling limit\footnote{
It is important to stress that this is only a technical and not a conceptual
problem. In fact, for instance, for the
simplest possible gauge theory, i.e. the $Z_2$ gauge theory in three
dimensions, the SC expansion has been
pushed to so high levels~\cite{arisue}
 that it gives results for the lowest masses of the
spectrum which are comparable in precision with those  obtained with MC
simulations~\cite{z2}.}.

A second reason of interest, which is particularly important from the point of
view of the comparison that we have in mind in these lectures, is that in the
framework of SC
expansions  a string description of LGT arises in a very natural way. In fact
both the partition function and the correlators of gauge invariant operators
become in the SC limit {\sl sums over suitably chosen surfaces}.
  This is
to be contrasted with the case  of ordinary (not gauge
invariant) field theories regularized on the lattice where the SC expansion 
becomes a sum over paths instead of surfaces. 

In order to clarify the above statements
let us study in more detail how the strong coupling expansion works
 in LGT.  
 We need first of all to expand the
Wilson action in the character basis of $SU(N)$. This is a standard problem in
group theory and we shall discuss it in the exercise 3 below. The next step
consists in substituting this expansion in each plaquette and then perform the
group integrations over the links. One easily sees that, 
due to the orthogonality
relations of characters, the only terms which survive in the expansions are
those in which the plaquettes are ``glued'' together to form a closed surface.
If we are interested in the partition function this is the end of the story. 
The partition function becomes a  weighted
 sum over all
possible closed surfaces that we can construct on the lattice.
 As anticipated above, this sum strongly resembles the
 discretized version of some (unknown) string-type theory. If we are instead
interested in the expectation value of some
 gauge invariant operator described by a closed contour $\Gamma$
 it is easy to see that the first contribution in the strong coupling limit is
 given by the minimal surface bounded by $\Gamma$. Further terms in the
 expansion will come from the fluctuations around this minimal surface. Again,
 this result strongly suggests a string like description for these observables.

As an example we report in the exercises 4 and 5 the calculation of the first 
term in the SC expansion for the Wilson loop and
for the lowest glueball. The results are (see eq.s (\ref{esig})
 and (\ref{egb})):
\eq
\sigma=-\log D_f(\beta)
\en
\eq
M(0^{++})=-4\log D_f(\beta)
\en
where $f$ denotes the fundamental representation and 
$D_f(\beta)$ is given, for a
generic value of $N$ by eq.(\ref{dierre})

Let us see two examples which are particularly relevant for us: 
the $SU(2)$ case which is
the simplest possible non abelian $YM$ theory and the large $N$ limit which is
the limit in which the results obtained using the AdS/CFT correspondence
 are expected to hold.

For $SU(2)$ we have (see eq.(\ref{esu2})
\eq
D_f(\beta)~\equiv~ D_\frac12(\beta)~=~\frac{I_2(\beta)}{I_1(\beta)}
\en
where $I_1$ and $I_2$ are modified Bessel functions of integer order.

In the large $N$ limit we find first of all 
that a consistent limit can only be obtained by sending also $\beta\to\infty$
and
keeping $\beta/ N$ fixed (in agreement with the 't Hooft prescription that we
shall discuss in sect.~\ref{s3.4}). In this limit we find 
(see eq.(\ref{elargeN}))
\eq
D_f(\beta/N)=\frac{\beta}{2N}~.
\en

The discussion of this section only gives a very short account of all the
richness and complexity of SC expansions in LGT. 
The standard reference for further readings
 is~\cite{dz} where a very detailed and complete discussion of the subject
 can be found. 

\begin{itemize}
\item {\bf Exercise 3: Character expansion for the $SU(N)$ group.}

Construct the character expansion of the Wilson action eq.(\ref{w1}).
\item{\bf Exercise 4: Evaluate the first  order in the strong 
coupling expansion of the Wilson loop in $SU(2)$ $YM$ theory.}
\item{\bf Exercise 5: Evaluate the first  order in the strong 
coupling expansion of the lowest glueball mass in $YM$ theories.}

\end{itemize}

\subsection{Scaling.}
\label{s3.3}
Once we have obtained with some non-perturbative method the value of a
dimensional physical quantity for some fixed value of $\beta$ we face the
problem of extracting a continuum limit estimate out of these numbers.
To this end one must first
 check that the values that we have measure
 scale as a function of $\beta$ according to the
 expected
asymptotic scaling behaviour. However it is often much simpler to test the
behaviour of adimensional ratios of different observables. The reason is that
very often the deviations from the asymptotic
scaling behaviour (due for instance to irrelevant operators) cancel in the
ratio. As a general rule the adimensional ratios are more stable than the
single observables. 

Notwithstanding this trick, 
one has in general to face rather large deviations from
the expected scaling behaviours.  The obvious solution to this problem would be
to study very large values of $\beta$. However both SC expansions and MC
simulations cannot be easily pushed upto these regions. SC expansions are centered
in $\beta=0$ and very high orders are needed to obtain stable results at large
$\beta$. MC simulations suffer of the so called ``slowing down'' problem. As
the correlation length increases it becomes more and more difficult to obtain
statistically independent configurations. Thus, practically, MC simulations are
confined to not too large values of $\beta$. It is thus very important to have
a good control of the systematic (not statistical!!) errors involved in 
extrapolating toward the continuum limit the MC results. There is by now 
a well developed technology to play this game. However one should always
consider the results obtained from MC simulations with some caution.

A completely different problem is represented by the possible presence of
phase transitions in the phase diagram of the model. If the range of $\beta$
values that we can study is separated from the continuum limit $\beta=\infty$
by a phase transition (which cannot be overcome by a suitable modification of
the action), then there is no hope to be able to obtain reliable
continuum estimates of the physical quantities. 

The most important example of such a situation is represented by the SC 
expansion
for the string tension. For a finite value $\beta_r$ of the coupling the
Wilson loop undergoes a phase transition (the well known roughening transition)
which does not allow to extend the SC series up to $\beta=\infty$. For this
reason in studying the string tension we must only resort on MC simulations.

\subsection{Large $N$ limit and the loop equations.}
\label{s3.4}
We have seen in the previous sections that the main advantage of the lattice
discretization is that it is a truly non-perturbative regularization of QCD.
The price that one has to pay is the introduction of the lattice spacing and
the difficult part of the game becomes the elimination of this new scale so as
to reach the correct continuum limit of the theory. It would be of great
importance to have some kind of non-perturbative insight of the theory
directly in the continuum. The large $N$ limit of 't Hooft~\cite{h74}
 represents the most concrete proposal in this direction. 

't Hooft's proposal starts from the observation that
in non-abelian gauge theories another dimensionless quantity exists besides the
bare coupling constant $g$. It is the number of colours $N$.

The main problem with QCD is that
 $g^2$ is not a good expansion parameter for the theory
 since (as we have seen in sect.~\ref{s2.3}) it runs with the cutoff.
 In fact the
 correct way to deal with $g^2$ is to trade it and the cutoff for the
 Renormalization Group invariant scale $\Lambda_{QCD}$ (see eq.(\ref{cs0})).
't Hooft was able to show that  $1/N$ is indeed a much better
 expansion parameter than $g^2$ and that in the large $N$ limit the theory
 drastically simplifies. Before discussing these simplifications let us
 concentrate on the limit itself.

If we look at eq.(\ref{cs}) we see that $g^2$ always appears multiplied by
$\beta_0$ which is proportional to $N$, thus if we want to keep $\Lambda_{QCD}$
fixed as we take the $N\to\infty$ limit we must simultaneously take $g\to0$
keeping the so called 't Hooft coupling $\lambda\equiv g^2N$ fixed.
In this limit the Feynman graphs are well defined and can be organized in a
perturbative expansion in powers of $1/N$. Now the seminal observation 
of `t Hooft was
that this perturbative expansion is actually an expansion in the $topology$ of
the Feynman graphs. 
To understand this statement let us first of all explain what do we
mean for ``topology of a graph'' .

 Let us assume the simplest possible definition of graph, i.e. a collection 
$(V,E)$ of Edges $E$ and Vertices $V$ in which all 
the edges are of the same type. Then it is possible to associate a genus to
each graph by noticing that each graph can be unambiguously embedded
 in a 2d
 Riemann surface and hence can be characterized by its genus. For instance the
 graphs with genus 0 are the planar graphs which, in fact, are exactly
 those graphs  which can be ``drawn'' on a 
 sphere. It is far from obvious that the Feynman graphs of a non-abelian gauge
 theory (with different propagators for quarks and gluons) fall into the above
 definition, but 't Hooft was able to give a set of recipes to allow this
 identification.

  Let us now study as an example the $1/N$ expansion of the
 free energy. It turns out that the first term in the expansion is proportional
 to $N^2$ (this was to be expected since if we keep $\lambda$ fixed then the
 Lagrangian itself becomes of order $N^2$) and that only even powers of $1/N$
 appear. Thus the expansion can be written as:
\eq
F=\sum_{{\bf g}=0}^{\infty}N^{2-2{\bf g}}F_{\bf g}(\lambda)
\en
where the $F_{\bf g}$ are complicated unknown functions of $\lambda$ which,
 in QCD-like theories are better expressed as functions of $\Lambda_{QCD}$.

The remarkable result of 't Hooft is that $F_{\bf g}(\lambda)$ 
only contains Feynman
graphs of genus ${\bf g}$. 
Thus the expansion obtained in this way strongly resembles 
the loop expansion in string theory if we identify $N$ with the inverse string 
coupling $1/g_s$. This is one of the strongest indications in favour of a
string-like description of QCD.

 The above observations have some important consequences. Let us discuss them
in detail.
\begin{itemize}
\item {\bf The planar limit}

In the large $N$ limit only the planar graphs survive (${\bf g}=0$) 
and the original
theory greatly simplifies.
In two dimensions the planar theory can be solved exactly and thus, at least 
in this case, an exact solution for QCD in the large limit can be obtained. 

\item {\bf The Master Field.}

The most interesting implication of the large $N$ limit is the idea of the so
called "master field". The starting point is the observation 
 that in the large $N$ limit disconnected diagrams are in general dominating.
 This implies that if we study the vacuum expectation value of a collection of
 (gauge invariant) operators ${\cal O}_i$, all the propagators (i.e. connected
 correlators) joining together two of the operators disappear in the large $N$
 limit and the VEV becomes the product of the VEV's of the single operators.
\eq
\vev{{\cal O}_1{\cal O}_2\cdots {\cal O}_n}=
\vev{{\cal O}_1}\vev{{\cal O}_2}\cdots\vev{{\cal O}_n}
\en
This means that
in the large $N$ limit the functional integral defining the above correlation
function must be dominated by a single field configuration, which is usually
called {\sl master field}~\cite{master}. 

This important result gave the hope that it could be possible to explicitly
 solve QCD in the large $N$
 limit and was the starting point of a large number of
 papers discussing the peculiar properties of the master field and the
 equations which it must satisfy. 

These methods were successfully applied to
 some simple problems for which indeed a master field could be
 explicitly found.
However  in the most interesting cases of QCD in
three and four dimensions none of these approaches
 has led so far to an explicit expression for the master field. One of the 
reasons of interest in the AdS/CFT correspondence is that it gives the first
 example of a master field solution for a set of non trivial, interacting,
 gauge theories in more than two dimensions.

\item {\bf The loop equations.}

A related result is that in the large $N$ limit the Wilson loop satisfies a
closed set of equations, which are called ``loop equations"~\cite{migmak}.
These equations can be derived in a rigorous way
 in the framework of the lattice  regularization and it can be shown that
they are solved by the master field of the theory.
 They can also be derived, at
 least formally, for the continuum theory where it can be shown that they are
 formally satisfied order by order in the perturbative expansion (for a review
 see~\cite{rev_loop}). 

 Unfortunately, despite many efforts,
 these equation could be solved
 explicitly only in the  case of 2d $YM$ theories~\cite{kk80} 
 and it turned out to be impossible to extend this solution also
 to the $d>2$ case. However, even if they cannot be solved exactly, these
 equations are a very interesting object themselves. In fact
 they are, so to speak, an intrinsic, defining, property of $YM$ theories. 
 In the lattice version of the theory the
 loop equations hold for any value of the coupling, both near the continuum
 limit and deep in the strong coupling region. Thus they are a perfect tool
 to test also in the strong coupling regime if a theory which we hope could be
 identified with QCD displays the correct large $N$ behaviour.

\item{\bf Eguchi-Kawai models.}

The most remarkable feature of the lattice  version of the loop equations is
that they allow in the large $N$ limit to
reduce the whole lattice model to a much simpler one plaquette model, while 
keeping the full physical content of the original model. 
This idea was first proposed  by  Eguchi and Kawai \cite{ek} and subsequently 
perfected by several authors and it is based on the observation that in
the large $N$ limit a suitably twisted\footnote{The twist 
consists in a suitable phase factor belonging to the center of SU$(N)$ 
that multiplies each plaquette variable in the action.} 
 lattice gauge theory  on a lattice 
consisting of just one site and one link variable for each
space-time direction generates the same set of loop equations as a theory 
defined on a large lattice, typically consisting of $N^{(d+1)/2}$ sites.
Hence twisted one plaquette models
can be used to describe lattice gauge theories on large lattices, 
by essentially mapping space-time degrees of freedom into internal 
degrees of freedom. A general review of the Eguchi--Kawai model, 
 can be found in~\cite{das}. 

Let us conclude by noticing
that in the framework of the AdS/CFT correspondence
in order to obtain  non-supersymmetric $YM$-like theories, one must look at the
finite temperature behaviour of the large $N$ model in which the ``time''
direction is compactified. Trying to perform the same construction 
in  Eguchi--Kawai type models turns out to be a rather non-trivial task due to
the interplay between the twists needed to define the EK model and those
induced by the periodic boundary conditions in the time direction.
We refer the reader to~\cite{bcdp}
 for a discussion of this problem and a review of the remarkable
properties of the large $N$ limit in finite temperature LGTs.
 
\end{itemize}

\subsection{The effective string picture.}
\label{s3.5}

\subsubsection{ The roughening transition}
\label{s3.5.1}

It is important to stress that the roughening transition is $not$ a phase
transition of the model itself. At the roughening point the
LGT partition function is regular, and  the correlation length of the model
(the inverse of the lowest glueball mass) is finite. 
The roughening point is instead the point in which the expectation value
of  one particular observable: the Wilson loop becomes singular.
This means that for all the observables different from the Wilson loop (and in
particular for instance for those related to the glueball states) there is no
obstruction (i.e. no phase transition in between) 
to reach the continuum limit starting
from the strong coupling phase.

 On the contrary, as far as the Wilson loop is
concerned,
the confining regime of LGTs contains (in general) two phases: the 
strong coupling phase and the rough
phase. The two are separated by the roughening transition 
which is the point in which the  strong coupling expansion of the Wilson loop
ceases to converge~\cite{rough,lsw}.
 These two phases are related to
two different behaviors of the quantum fluctuations of the 
flux tube around its
equilibrium position~\cite{lsw}. 
In the strong coupling phase, these fluctuations are
massive, while in the rough phase they become massless 
and hence survive in the
continuum limit. The inverse of the mass scale of these fluctuations
 (which is completely different
from the glueball mass scale
 and only appears in the model if we study the expectation
value of the Wilson loop) can be considered as a new 
correlation length of the model. It is exactly this new correlation length
 which goes to infinity at the roughening point and
determine the singular behaviour of the Wilson loop.  This fact has several
consequences:
\begin{description}
\item{(a)} The flux--tube fluctuations can be described by a suitable
two-dimensional massless quantum field theory,
where the fields describe the
transverse displacements of the flux tube. This quantum field theory is 
expected to be very
complicated and will contain in general non renormalizable interaction terms
\cite{lsw,cfghpv}. 
However, exactly because these interactions are non-renormalizable, their
contribution becomes negligible in the infrared limit (namely for large Wilson
loops). In this infrared limit the QFT becomes a conformal invariant 
field theory (CFT)
(See {\it e.g.} chapter 9 of Ref.~\cite{idbook}
 for a comprehensive review on  CFTs).
\item{(b)} The massless quantum fluctuations delocalize the flux tube
 which acquires a nonzero width, which diverges logarithmically as the
 interquark distance increases~\cite{lmw, width}.
\item{(c)} The quantum fluctuations
 give a non-zero contribution to the interquark potential,
which is related to the partition function of the above $2d$ QFT.
Hence if the $2d$ QFT is simple enough to be exactly solvable (and this is in
general the case for the CFT in the infrared limit) also these
contributions can be evaluated exactly.
\item{(d)} In the simplest case, this CFT  is simply
the two dimensional conformal field theory 
of $(d-2)$ free bosons ($d$ being the number of spacetime dimensions 
of the original gauge model); its exact solution will be discussed in 
exercise 6.

\end{description}

\vskip 0.5cm
\subsubsection{ Finite Size Effects: the L\"uscher term.}
\label{s3.5.2}

\vskip 0.1cm
The feature of the effective string description which is best
 suited to be studied by numerical methods is the presence of 
finite--size effects. 

Wilson loops in the confining phase are classically expected to obey the 
area law (see eq.(\ref{area})).
This law is indeed very well 
verified in the strong coupling regime (before the
roughening transition), but it is  inadequate to describe the Wilson
loop in the rough phase. In this phase the strong coupling expression
 must be  multiplied by the 
partition function of the $2d$ QFT describing  the quantum
fluctuations of the flux tube. This QFT in the infrared limit becomes a 2d CFT
whose
 partition function   $Z_{q}(R,T)$
 can be in some cases 
 evaluated exactly. We shall discuss in exercise~6 an example of this type of
 calculations. 

Eq.~(\ref{area}) in the rough phase becomes:
\be
<W(R,T)>=e^{-\sigma RT+p(R+T)+k}Z_{q}(R,T)~.
\label{quantum}
\ee

In general, even if one cannot give the exact expression of 
 $Z_{q}(R,T)$ it is always possible to express its dominant contribution to the
interquark potential, (i.e. in the limit $T>>R$) which turns out to be:

\be
\lim_{T\to\infty} \frac{1}{T}\log Z_{q}(R,T)= \frac{c\pi}{24 R}~~,
\label{limit}
\ee
where $c$ is the central charge of the CFT. In the simplest possible case,
namely when the CFT describes a collection of $n$ free bosonic fields, 
we have
$c=n$. Thus for the free boson realization of the effective string 
theory, 
we find $c=d-2$. This is the result obtained by
L\"uscher, Symanzik and Weisz in~\cite{lsw}.

The interquark potential is thus given (neglecting an irrelevant constant) by:
\be
V(R)=-\lim_{T\to\infty} \frac{1}{T}\log\Br W(R,T)\kt
=\sigma R - \frac{c\pi}{24 R}~~.
\label{luescher}
\ee

The $1/R$ term in the potential is the finite size effect mentioned above; 
it is
completely due to the quantum fluctuations of the flux tube and, if
unambiguously detected, it represents a strong evidence 
(the strongest we have) 
in favor of the effective string picture discussed above. Moreover if the 
measurement
is precise enough we can in principle extract numerically the value of $c$ and
thus select which kind of effective string model describes the infrared 
regime of the 
LGT under examination.

Unfortunately, if one tries to evaluate the $1/R$ contribution in $SU(2)$ or
$SU(3)$ gauge theories in (3+1) dimensions one faces a non trivial problem.
In LGTs in (3+1) dimensions 
with continuous gauge groups the interquark potential
 has  another contribution of $1/R$ type which has a completely different
 origin. It is due  to the one gluon exchange. It can be
 evaluated perturbatively, and it
 exists only in the ultraviolet regime, namely for
 small Wilson loops.  
 Even if it holds only in the perturbative regime, we cannot
fix a sharp threshold after which it disappears, so it could well be that,
 in the
set of large Wilson loops from which we extract our data
we find
a superposition of the two terms. There are two ways
to avoid this problem:
\begin{itemize}
 \item
  Study LGT in three dimensions where the perturbative term has 
a $\log R$ form
  instead of $1/R$, and does not mix up with the string contribution.
 \item
Study Wilson loops with comparable values of $T$ and $R$. In this case, 
  the whole
functional form of the two interaction terms becomes important. These are
completely different and thus can be separated.
\end{itemize}

Since the beginning of eighties several numerical works have been done to study
this problem. The main results can be summarized as follows:

 \begin{description}
  \item{(a)} A $1/R$ term 
exists in the potential. In the case of (3+1) LGT with
  continuous gauge group it can be observed also at very large distances, thus
  it is unlikely that it can be only due to the one gluon exchange.

  \item{(b)} A similar $1/R$ term has been observed in various (2+1) models.
In these cases the string interpretation is unambiguous.

  \item{(c)} The same correction is found in very different LGTs, ranging from
  the $3d$ Ising gauge model to the $4d$ $SU(3)$ model. 
This remarkable universality
  is an important feature of these finite size effects of the effective string
  description.

  \item{(d)} The central charge has been measured with rather good
  precision. The numbers are in good agreement with the
  $c=d-2$ prediction of L\"uscher Symanzik and Weisz for the (2+1) dimensional
  theories. They slightly differ in the (3+1) case. It is well possible that
  this is only due to the superposition of Coulomb potential.

  \item{(e)} In the case of the simplest possible gauge theory, i.e. the gauge
  Ising model in 3 dimensions a high precision test of eq.(\ref{quantum}) has
  been performed~\cite{cfghp}. Not only the central charge, but the 
  whole functional form of the $Z_q$ correction was tested and full
  agreement with the effective string predictions was found.

\end{description}

\subsubsection{ String Universality}
\label{s3.5.3}
We have seen (point (c) of the previous section) that the same effective string
corrections have been found in all the LGT which have been studied up to now.
As a matter of fact not only the string corrections, but also other 
features of the  infrared regime of
LGTs  in the confining phase  display a high
degree of universality, namely they seem not to
 depend on the choice of the gauge group. This is the case for instance of the
 ratio between the critical temperature and the square root of the string
 tension, or the behavior of the spatial string tension above the deconfinement
 transition.  
All these examples show a substantial independence on the gauge group and a
small and smooth dependence on the number of spacetime dimensions.

This ``experimental fact''
 has a natural explanation in the context of an effective string model:
even if in principle different gauge models could be described
 by different string theories,
 in the infrared  regime, as the interquark distance increases 
all these different string theories 
flow toward the common fixed point which is not anomalous and corresponds, 
in the simplest case discussed above,
to the two dimensional conformal field theory 
of $(d-2)$ free bosons.
Also the small dependence on the number of spacetime 
dimensions of the theory is
well predicted by the effective string theory.

It is well possible that this string universality is only due to the fact that
we are addressing with our simulations the simplest possible gauge theories and
that looking to more complicated models a whole spectrum of effective
string theories could appear, similarly to what happens in standard 2d 
conformal field theories. However it is interesting in this respect to notice
that by resorting only to
 the  basic  property of Osterwalder-Schrader positivity (which must be true
 in the most general unitary LGT)
one can obtain~\cite{bachas} a constraint on the 
possible values of the coefficient of the $1/R$
correction (and hence of the central charge of the effective string theory).
It turns out that the interquark potential must be both monotonic increasing
and concave~\cite{bachas}
 thus implying that the central charge of the effective string must
be non-negative. We shall come back to this observation when discussing the
AdS/CFT results for the interquark potential.

\begin{itemize}
\item{\bf Exercise 6: The effective string contribution to a
rectangular Wilson loop.}
 
 Construct the effective string contribution to a
rectangular Wilson loop (the $Z_q(R,T)$ term in eq.(\ref{quantum})) assuming a
 simple Nambu-Goto action for the string.
\end{itemize}

\subsection{The string tension.}
\label{s3.6}

\subsubsection{(3+1) dimensions}
\label{s3.6.1}
\begin{figure}[htb]
\centerline{\epsfxsize=10truecm\epsffile{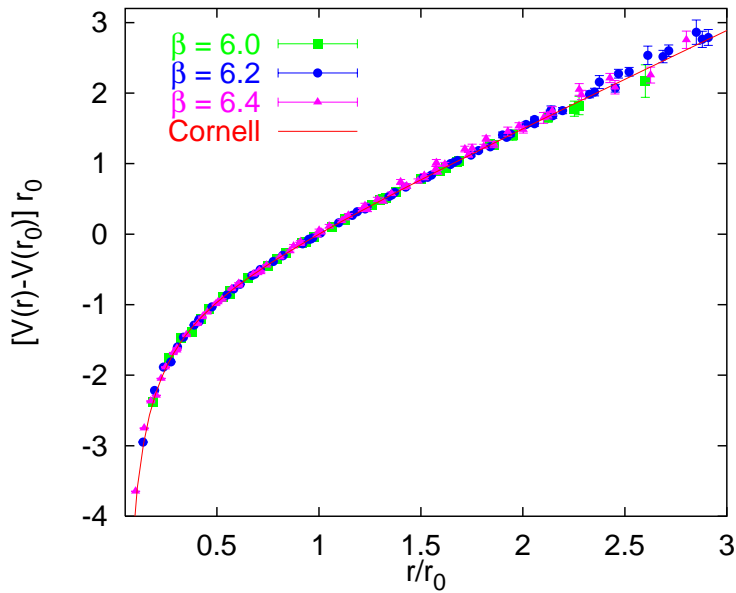}}
\caption{The quenched Wilson action $SU(3)$ potential in (3+1) dimensions,
 normalized to $V(r_0)=0$.}
\label{ads_pot}
\end{figure}

The best way to discuss the present status of the lattice results on the
interquark potential is to look at fig.~\ref{ads_pot} where the interquark
potential for the (3+1) dimensional $SU(3)$ model in the quenched approximation
is displayed. 

The figure is taken from~\cite{bali} (to which we refer for a thorough
discussion of the interquark potential) and is a a compilation of data reported
in Refs~\cite{Bali:1992ab,Bali:1997am}. 

Let us briefly comment this figure. This will also give us the
opportunity to explain how LGT results are usually presented in the literature.

\begin{itemize}

\item
Both the potential and the interquark distance are measured in units of $r_0$. 
This scale is obtained by looking at
the intermediate range in the interquark potential.
While the large-$r$ part of the potential is characterized by the string tension
$\sigma$,
one can characterize its behaviour at intermediate
distances by the distance $r_0$ at which the force, $F$, has a 
particular value. It has become customary to use the
particular definition $r_0^2F(r_0)=1.65$ (which corresponds
to a value that can be calculated with precision on 
the lattice and which can be estimated with some reliability
from the observed spectrum of heavy quark systems). In physical units
this corresponds to $r_0\approx 0.5$~fm.
\item
The important consequence of this choice
 is that in this way all the physical quantities ($r$
and $V(r)$) are measured in physical units and not in terms of the lattice
spacing. The whole complexity of the scaling function eq.(\ref{cs}) is hidden
in $r_0$ and the two combinations $r/r_0$ and $V(r) r_0$ are adimensional
ratios, in the sense discussed in sect.~\ref{s3.3}. They are renormalization
group invariant quantities and must keep the same value as the cutoff is
changed (or, that is the same, as $\beta$ is changed) if we are in the scaling
region. Thus we have an immediate and very effective test of scaling: data taken
at different values of $\beta$ must overlap in the figure.

\item 
This is indeed the case for the data reported in the figure which
 correspond to three samples of data
(denoted by squares, triangles and circles respectively), 
obtained with MC simulations
 performed at three different values of $\beta$ (see the inset in the figure).
The perfect overlap of the data is telling us that, at least for this
observable,
the scaling region is reached already at $\beta=6.0$.

\item 
By using the scaling function (and the value $r_0\approx 0.5$~fm) we may obtain
the value, in physical units, of the lattice spacing for the three samples in
the figures. They correspond to $a\approx 0.094$~fm,
0.069~fm and 0.051~fm, respectively. This gives an idea of the size of the
``grid'' of our lattice approximation. 

\item
Looking at the figure we see that the
maximum interquark distance that we one can study is about 1.5~fm (recall 
that we are in the quenched approximation, so the interquark string
cannot break).
If we tried to push the quark and antiquark pair further apart we would have
to face two types of problems. First we would have to fight against increasing 
statistical errors (denoted in the figure by the errors bars)
due to the fact that as the Wilson loops become larger and larger, since they
are exponentially depressed due to the area law, the signal to noise ratio
becomes smaller and smaller and too long runs are needed to obtain statistically
significant results. Second, one must take into account the systematic errors
due to the finite size of the lattice in which the Wilson loops are immersed. 
The lattice size must be much larger than the Wilson loop size to allow one to
neglect these systematic errors, but again larger lattices require much more
time to obtain statistical independent configurations.

\item
The data are fitted with the so called ``Cornell potential'', which is
essentially eq.(\ref{luescher}) 
in which the coefficient of the $1/R$ term is kept
as a free parameter:
\eq
V(r)=V_{self} +\sigma r- \frac{e}{r}
\en
The result of the two parameter 
fit\footnote{The additive self-energy contribution,
(associated with the perimeter term in the area law)
 is eliminated from the fit by the parametrisation-independent normalization of
the data to $V(r_0)=0$.} is plotted in the figure as a continuous
line. One can directly see that the data agree very well with the proposed
function. The best fit value for $e$ is  $e=0.295$ which is slightly higher than
the bosonic string prediction. This could be due to the interplay with the one
gluon exchange contribution or to the fact that in the Cornell
 approximation one is
neglecting the subleading (log type) contributions of the effective string.
However it could also be the signature that the
 effective string description of the model is more complicated than the simple
 free bosonic model. 

\item
From the fit we also obtain a best fit estimate for the string tension.
This is the value that we shall use in the next subsection as a scale 
 to measure the glueball masses. We report in tab.~\ref{gb3} the result for both
 $SU(2)$ and $SU(3)$ in units of $\Lambda_{\overline{MS}}$.
 We also report in the same table for comparison the string
 tension in units of $r_0$ and $T_c$ (the deconfinement temperature).

\item
Looking carefully at the figure one can see that at small
distances the data points lie somewhat
above the curve, indicating a weakening of the
effective coupling. This is a signature of the onset of
 asymptotic freedom at short distances.

\end{itemize}

\begin{table}
\begin{center}
\begin{tabular}{|r|c|c|}\hline
     & $SU(2)$ & $SU(3)$ \\ \hline \hline
$r_0 \sqrt\sigma$   & $1.201\pm0.055$ & $1.195\pm0.010$ \\ \hline
$T_c/\sqrt\sigma$   & $0.694\pm0.018$ & $0.640\pm0.015$ \\ \hline
$\Lambda_{\overline{MS}}/\sqrt\sigma$   &  & $0.531\pm0.045$ \\ \hline
\end{tabular}
\caption{\label{gb3} Continuum  limit value of the adimensional ratios
of the string tension with the scale $r_0$, the deconfining
temperature, $T_c$
 and the (zero-flavour) $\Lambda$ parameter in
the ${\overline{MS}}$ scheme for the $SU(2)$ and $SU(3)$ $YM$ theories in (3+1)
dimensions.} 
\end{center}
\end{table}

The next step is now to study the large $N$ limit of the string tension. We
shall address this problem in the simpler case of (2+1) dimensional theories

\subsubsection{(2+1) dimensions}
\label{s3.6.2}
The analysis is similar to that discussed in the previous section, but in
 this case it is possible to perform simulations also for larger values of $N$
in particular, in~\cite{teper1} results for $N=4$ and $N=5$ were obtained.
 It turns out
that these values are already large enough to perform a reliable large $N$ 
limit. 
Recall that in this case the coupling constant has the dimensions of a mass and
thus can be used to set the mass scale of the whole theory. 

It is thus natural in this
case to express the string tension in units of $g^2$. The results for the various
groups are:
\begin{equation}
{ {\sqrt\sigma} \over {g^2} } =
\left\{ \begin{array}{ll}
0.3353(18) & \ \ \ \mbox{SU(2)} \\
0.5530(20) & \ \ \ \mbox{SU(3)} \\
0.7581(40) & \ \ \ \mbox{SU(4)} \\
0.9657(54) & \ \ \ \mbox{SU(5)}
\end{array}
\right.
\label{d=3}
\end{equation}

It is easy to see that these values increase linearly as a function of $N$. This
agrees with the discussion of sect.~\ref{s3.4} on the 
large $N$ limit where it was
shown that the natural coupling in this limit is the 't Hooft combination
$g^2N$. 

If we try to fit the data of eq.(\ref{d=3}) keeping also into account the first
subleading term in $1/N$ we  see that it is
proportional to $1/N^2$. This too is a prediction of the large $N$ analysis
which is perfectly confirmed by the simulations.

The fit with two free parameters to the equation:
\begin{equation}
{ {\sqrt\sigma} \over {g^2 N} } = 
c_0 + {{c_1} \over {N^{2}}}
\label{7.1}
\end{equation}

gives as a result:
\eq
c_0=0.1975(10)~~,~~~~~~~~~~~~~~~~~
c_1=-0.119(8)~~~.
\en
with a very good confidence level (see~\cite{teper1} for the details). 
The value of $c_0$ obtained in this way represents the 
first example of a non-perturbative result in the large $N$ limit $SU(N)$ gauge
theories.

Let us conclude with two comments on this result
\begin{itemize}
\item
The fit gives an acceptable confidence level even if the $SU(2)$ result is taken
into account, this means that the large $N$ limit analysis (taking into account
also the first $1/N^2$ correction) holds all the way down to $N=2$

\item 
The fact that the data show the correct $N$ dependence
 is a highly non trivial test of 't Hooft
analysis, since it comes from a truly non-perturbative regularization and is
completely independent from the weak coupling arguments of sect.~\ref{s3.4}

\end{itemize}

\subsubsection{The space-like string tension in finite temperature LGT.}
\label{s3.6.3}

It is important to stress that the results discussed in the two previous
sections
strictly refer to the zero-temperature version of $SU(N)$ LGT.
In finite temperature LGT the interquark potential is obtained from the
connected correlator of Polyakov loops. Recall that in FTLGT the lattice is
asymmetric and we can distinguish between time-like and space-like Wilson
loops. The spacelike Wilson loops are those orthogonal to the compactified time
direction. If
 periodic boundary conditions in the time directions are imposed a timelike
 Wilson loop whose length in the $T$ directions is larger that the lattice size
naturally becomes a pair of Polyakov loops, and this is the true order parameter
for confinement. On the contrary the spacelike string tension is no longer an
order parameter of the theory. In general it is different from zero even in the
deconfined phase and (contrary to the naive expectation) 
increases as the temperature is increased! (for a
discussion of this issue see for instance~\cite{spacelike}). 
It will be important to remember this fact when we shall look at the 
Wilson loops in the framework of the AdS/CFT correspondence
 for non-supersymmetric
gauge theories. It will turn out that they are actually spacelike Wilson loops
of the underlying supersymmetric gauge theories. This explains why, while the
supersymmetric theories, being conformally invariant, are not confining the
spacelike Wilson loops (which will be interpreted as ordinary Wilson loops of
the non-supersymmetric theory) are indeed confining.

\subsection{The glueball spectrum.}
\label{s3.7}

\subsubsection{(3+1) dimensions.}
\label{s3.7.1}

In this section we summarize our present knowledge of the glueball spectrum in
(3+1) dimensions in the quenched approximation 
from Montecarlo simulations. The quantum numbers are presented with the standard
convention $J^{PC}$ while the asterisks refer to the radial excitations.
 The most precise results have been
obtained  in the case of
the $SU(2)$ and $SU(3)$ groups.
 They are reported in tab.~\ref{gb1} for $SU(2)$ and  in tab.~\ref{gb2} for 
$SU(3)$. We take this opportunity to show some of the ways in which these data
are usually presented in the lattice literature. In  tab.~\ref{gb1} the glueball
masses for the $SU(2)$ model are reported in units of the
string tension (second column) and  in units of the lowest glueball (last
column).
Notice the absence of any $C=-$ states in the $SU(2)$ case.
These values are taken from ref.~\cite{teper2} and the quoted errors take care
both of the statistical  and the systematic uncertainties.

\begin{figure}[thb]
\centerline{\epsfxsize=10truecm\epsfbox{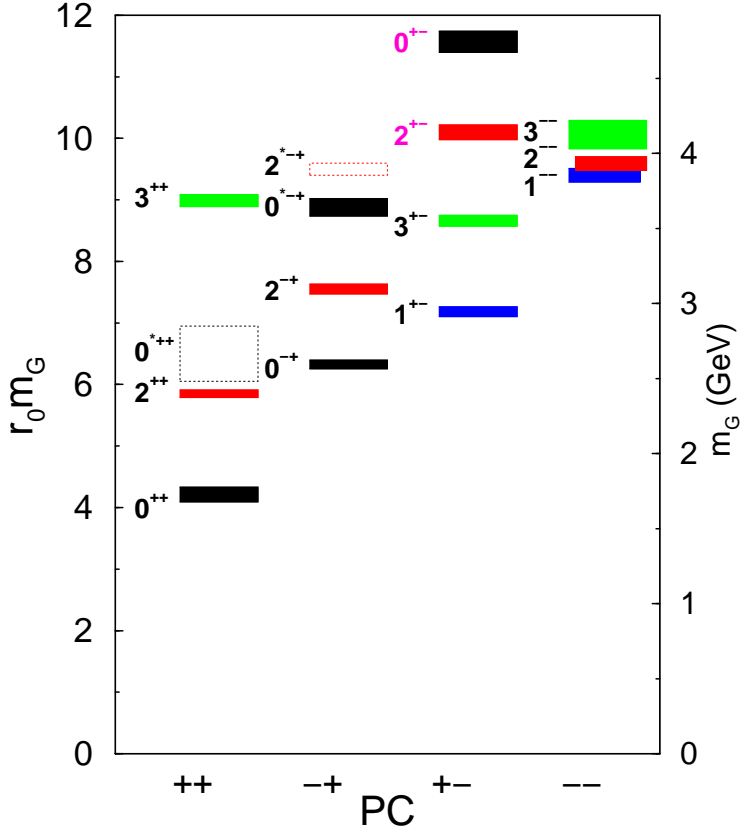}}
\caption{The glueball spectrum of $SU(3)$ gauge theory in (3+1) dimensions.}
\label{glueballs}
\end{figure}

In  tab.~\ref{gb2} we report the  glueball spectrum for $SU(3)$ first (in the
second column) in units of the scale $r_0$ (see the
discussion in sect.~\ref{s3.6.1}) and then (last column) in physical units
(MeVs).
These values are taken from ref.~\cite{Morningstar:1999rf}. 
The $SU(3)$ data are also plotted in fig.~\ref{glueballs} 
(taken again from Ref.~\cite{Morningstar:1999rf}).
The width of the 
 states in the figure corresponds to the combined statistical and systematic 
 uncertainties of the estimates.

Let us stress an important  non-trivial feature of the spectrum. 
Contrary to the naive expectations the $2^{++}$ state has a mass 
lower than that of the $1^{-+}$. This is not an accident, it happens in all the
model studied up to now (both in (2+1) and (3+1) dimensions, both with
continuous and discrete gauge groups) and can be considered as
 a fingerprint of $YM$ theories. We shall come back to this point when 
 discussing the glueball spectrum in the AdS/CFT framework.

\begin{table}
\begin{center}
\begin{tabular}{|l|c|c|}\hline
$J^{PC}$ &$m_G/\sqrt\sigma$ & $m_G/m_{0^{++}}$ \\ \hline
$0^{++}$         & 3.74$\pm$0.12 &  \\
$2^{++}$         & 5.62$\pm$0.26 &  $1.46\pm0.09$\\
$0^{-+}$         & 6.53$\pm$0.56 & $1.78\pm0.24$ \\
$2^{-+}$         & 7.46$\pm$0.50 & $2.03\pm0.20$ \\
$1^{++}$         & 10.2$\pm$0.5  & $2.75\pm0.15$ \\
$1^{-+}$         & [10.4$\pm$0.7] & [$3.03\pm0.31$] \\
$3^{++}$         & 9.0$\pm$0.7    &  $2.46\pm0.23$\\
$3^{-+}$         & [9.8$\pm1.4$]  & [$2.91\pm0.47$] \\ \hline
\end{tabular}
\caption{\label{gb1}
Continuum
limit extrapolation for the glueball masses for the $SU(2)$ $YM$ theory 
 in (3+1) dimensions. In the second column the masses are listed
 in units of the string tension, in the last column
 in units of the lightest scalar glueball mass.  
 Values in brackets have been obtained by extrapolating
from only two lattice values and so should be treated with caution.}
\end{center}
\end{table}

\newcommand{\bb}{\hspace{1.25ex}}
\begin{table}[t]
\begin{center}
\begin{tabular}{|l|l|l|}
\hline
 $J^{PC}$ & \hspace{3ex} $r_0 m_G$ & \hspace{2ex} $m_G$ (MeV) \\
\hline
 $0^{++}$                  & \bb 4.21 (11)(4)            & 1730 (50)\bb(80)  \\
 $2^{++}$                  & \bb 5.85 (2)\bb(6)          & 2400 (25)\bb(120) \\
 $0^{-+}$                  & \bb 6.33 (7)\bb(6)          & 2590 (40)\bb(130) \\
 $0^{*++}$                 & \bb 6.50 (44)(7)$^\dagger$  & 2670 (180)(130)  \\
 $1^{+-}$                  & \bb 7.18 (4)\bb(7)          & 2940 (30)\bb(140) \\
 $2^{-+}$                  & \bb 7.55 (3)\bb(8)          & 3100 (30)\bb(150) \\
 $3^{+-}$                  & \bb 8.66 (4)\bb(9)          & 3550 (40)\bb(170) \\
 $0^{*-+}$                 & \bb 8.88 (11)(9)            & 3640 (60)\bb(180) \\
 $3^{++}$    & \bb 8.99 (4)\bb(9)          & 3690 (40)\bb(180) \\
 $1^{--}$    & \bb 9.40 (6)\bb(9)          & 3850 (50)\bb(190) \\
 $2^{*-+}$   & \bb 9.50 (4)\bb(9)  & 3890 (40)\bb(190) \\
 $2^{--}$    & \bb 9.59 (4)\bb(10)         & 3930 (40)\bb(190) \\
 $3^{--}$    &    10.06 (21)(10)           & 4130 (90)\bb(200) \\
 $2^{+-}$   &    10.10 (7)\bb(10)         & 4140 (50)\bb(200) \\
 $0^{+-}$   &    11.57 (12)(12)           & 4740 (70)\bb(230)  \\
\hline
\end{tabular}
\caption{\label{gb2}
Continuum
limit extrapolation for the glueball masses for the $SU(2)$ $YM$ theory 
 in (3+1) dimensions. In the second column the masses are listed
 in units of the $r_0$. The first error is the statistical uncertainty from the
 continuum-limit extrapolation and the second is an estimate of the systematic
 error due to the particular method (regularization on anisotropic lattices)
 used to evaluate the masses.
 In the last column the masses are reported in physical units.
 In this case the first error comes
 from the combined uncertainties in $r_0 m_G$, the second from the
 uncertainty in $r_0^{-1}=410(20)$ MeV.}
\end{center}
\end{table}

Let us now study
 the large $N$ limit of the glueball spectrum. As we did in the case of the
 string tension we shall first address the problem 
in the (2+1) dimensional case where everything is much
simpler and under control.
We shall then try in sect.~\ref{s3.7.3} below to perform the same analysis 
in the more interesting (3+1) dimensional case.

\begin{table}
\begin{center}
\begin{tabular}{|l|l|l|l|l|}\hline
\multicolumn{5}{|c|}{$m_G/\sqrt\sigma$} \\ \hline
state & $SU(2)$ &$ SU(3)$ & $SU(4)$ & $SU(5)$ \\ \hline
$0^{++}$         & 4.718(43) & 4.329(41) & 4.236(50) & 4.184(55) \\
$0^{++\ast}$     & 6.83(10)  & 6.52(9)  & 6.38(13) & 6.20(13) \\
$0^{++\ast\ast}$ & 8.15(15)  & 8.23(17) & 8.05(22) & 7.85(22) \\ 
$0^{--}$         &           & 6.48(9)  & 6.271(95) & 6.03(18) \\ 
$0^{--\ast}$     &           & 8.15(16) & 7.86(20) & 7.87(25) \\ 
$0^{--\ast\ast}$ &           & 9.81(26) & 9.21(30) & 9.51(41) \\
$0^{-+}$         & 9.95(32)  & 9.30(25) & 9.31(28) & 9.19(29) \\
$0^{+-}$         &           & 10.52(28) & 10.35(50) & 9.43(75) \\
$2^{++}$         & 7.82(14)  & 7.13(12) & 7.15(13) & 7.19(20) \\
$2^{++\ast}$     &           &          & 8.51(20) & 8.59(18) \\
$2^{-+}$         & 7.86(14)  & 7.36(11) & 6.86(18) & 7.18(16) \\
$2^{-+\ast}$     &           & 8.80(20) & 8.75(28) & 8.67(24) \\
$2^{--}$         &           & 8.75(17) & 8.22(32) & 8.24(21) \\
$2^{--\ast}$     &           & 10.31(27) & 9.91(41) & 9.79(45) \\
$2^{+-}$         &           & 8.38(21)  & 8.33(25) & 8.02(40) \\
$2^{+-\ast}$     &           & 10.51(30) & 10.64(60) & 9.97(55) \\
$1^{++}$         & 10.42(34) & 10.22(24) & 9.91(36) & 10.26(50) \\
$1^{-+}$         & 11.13(42) & 10.19(27) & 10.85(55)& 10.28(34) \\
$1^{--}$         &           & 9.86(23)  & 9.50(35) & 9.65(40) \\
$1^{+-}$         &           & 10.41(36) & 9.70(45) & 9.93(44) \\ \hline
\end{tabular}
\caption{\label{d3.1} Continuum limit values of the 
glueball masses for various (2+1) dimensional $SU(N)$ theories
 in units of the string tension.}
\end{center}
\end{table}

\begin{table}
\begin{center}
\begin{tabular}{|l|c|}\hline
state & $\lim_{N\to\infty} m/\sqrt\sigma$ \\
\hline 
$0^{++}$         & 4.065(55) \\
$0^{++\ast}$     & 6.18(13)  \\
$0^{++\ast\ast}$ & 7.99(22)  \\
$0^{-+}$         & 9.02(30)  \\
$2^{++}$         & 6.88(16)  \\
$2^{-+}$         & 6.89(21)  \\
$2^{-+\ast}$     & 8.62(38)  \\
$1^{++}$         & 9.98(25)  \\
$1^{-+}$         & 10.06(40) \\
\hline 
$0^{--}$         & 5.91(25) \\ 
$0^{--\ast}$     & 7.63(37)  \\
$0^{--\ast\ast}$ & 8.96(65)  \\
$0^{+-}$         & 9.47(116) \\
$2^{--}$         & 7.89(35)  \\
$2^{--\ast}$     & 9.46(66)  \\
$2^{+-}$         & 8.04(50)  \\
$2^{+-\ast}$     & 9.97(91)  \\
$1^{--}$         & 9.36(60)  \\
$1^{+-}$         & 9.43(75)  \\
\hline
\end{tabular}
\caption{\label{tab7}
The large $N$ limit of the mass spectrum in units of the
string tension in (2+1) dimension.}
\end{center}
\end{table}

\subsubsection{(2+1) dimensions.}
\label{s3.7.2}

We report in tab.~\ref{d3.1} the glueball spectrum in units of the square root of
the string tension for $N=2,3,4$ and $5$. The table is taken from~\cite{teper1}.
As anticipated above we observe again the inversion between the states of the
$J=2$ family and those of the $J=1$ one. 
The fact that in this case also $N=5$ data exist allows to make a reliable large
$N$ limit analysis.
Thus, following the discussion of sect.~\ref{s3.6.2} let us fit these data 
with
\begin{equation}
 {{M(J^{PC})}_N \over {g^2N}}
 = 
 M(J^{PC})_{\infty} + {{d_1} \over {N^2}}
\label{hh2}
\end{equation}
where we denote with $M(J^{PC})_N$ the mass of the glueball of quantum numbers
$J^{PC}$ in the $SU(N)$ theory.

For all the $J^{PC}$ values
 the fits turn out to have good confidence levels. The large $N$ limit
results are reported in tab.~\ref{tab7} (in units of the large $N$ string
tension). In the upper part of the table we have listed the glueball states
with $C=+$ for which also $SU(2)$ data exist. In the lower part of the table
we report 
the $C=-$ states which are obtained by fitting only data with $N>2$. 
 These are the numbers with which we shall compare the AdS/CFT
 predictions for (2+1) dimensions which we shall discuss in the next section.

\begin{table}
\begin{center}
\begin{tabular}{|l|c|c|c|}\hline
\multicolumn{4}{|c|}{$SU(4)$} \\ \hline
     & $\beta=10.7$ & $\beta=10.9$ & $\beta=11.1$  \\ \hline
$a\sqrt\sigma$     & $0.296\pm0.015$  & $0.228\pm0.007$  & 
$0.197\pm0.008$   \\
$am_{0^{++}}$      & $0.98\pm0.17$  & $0.75\pm0.07$  
& $0.77\pm0.06$   \\  
$am_{0^{++\star}}$ & $1.54\pm0.44$  & $1.39\pm0.16$  
& $1.03\pm0.14$   \\ 
$am_{2^{++}}$      & $1.28\pm0.27$  & $1.21\pm0.11$  
& $1.04\pm0.12$   \\ \hline  
\end{tabular}
\caption{\label{gb4}
$SU(4)$ masses and string tensions in(3+1) dimensions
on $10^4$, $12^4$ and $16^4$ lattices at $\beta=$10.7, 10.9
and 11.1 respectively.}
\end{center}
\end{table}

\begin{table}
\begin{center}
\begin{tabular}{|r|c|}\hline
\multicolumn{2}{|c|}{$SU(\infty)$} \\ \hline
$m_{0^{++}}/\sqrt\sigma$     &   $3.56\pm0.18$   \\
$m_{2^{++}}/\sqrt\sigma$     &   $4.81\pm0.35$   \\
$T_c/\sqrt\sigma$   &   $0.597\pm0.030$  \\
$r_0 \sqrt\sigma$   &   $1.190\pm0.043$  \\ \hline
\end{tabular}
\caption{\label{gb5}
Various adimensional ratios in (3+1) dimensions,
 after extrapolation to $N=\infty$
from $N=2$, $N=3$ and $N=4$;
$assuming$ the validity of eqn(\ref{C1}) all the
way down to $N=2$.}
\end{center}
\end{table}

\subsubsection{Large $N$ limit in 3+1 dimensions}
\label{s3.7.3}

In order to perform a reasonable large $N$ limit also in (3+1) dimensions we
need at least few informations also on the $SU(4)$ theory.
In Tab.~\ref{gb4} are reported some of the existing data for $SU(4)$ taken
from~\cite{teper2}. They deal only with the lowest three states of the
spectrum, but at least for them, they allow
 a tentative large N extrapolation.

In fact we see from tables~\ref{gb1},\ref{gb2} and~\ref{gb4}
that the physical  properties of $SU(2)$, $SU(3)$ and $SU(4)$ 
gauge theories are 
very similar. Assuming, as in the $2+1$ dimensional case that
we are already close to the $N = \infty$ 
limit even with $N=2,3,4$ we can 
extrapolate the glueball masses (in units of $\sqrt\sigma$)
 to the $N=\infty$ limit using
\begin{equation}
{m \over {\sqrt\sigma}}\Biggl|_{\!N}
=
{m \over {\sqrt\sigma}}\Biggl|_{\!\infty}
+
 \ {c\over{N^2}}
\label{C1}
\end{equation}
In this way  we obtain the values displayed in 
table~\ref{gb5} (see~\cite{teper2} for the details).
 While these results are slightly less stable than the (2+1)
dimensional ones they represent nevertheless the first non-perturbative results
in the large $N$ limit of $SU(N)$ gauge theories in (3+1) dimensions. As such they
are of the greatest importance. 
We shall use them to discuss the validity of the 
 AdS/CFT approach in the (3+1) dimensional case
 in the next section.

\newpage

\section{AdS/CFT.}
\label{s4}

As we mentioned in the introduction we shall assume in the following that the
reader is already acquainted with the theory behind the AdS/CFT correspondence.
 Some good reviews exist on the subject~\cite{rev_pdv,rev_agmoo} where the
 interested reader can find a thorough discussion of the correspondence
 and all the
 needed background material. The aim of this section is to provide
 the reader with
 the necessary
 information  to compare the physical picture
which emerges in the framework of the AdS/CFT correspondence
 with the results discussed in the previous sections following
 the lattice approach. For this reason we have organized this section in two
 parts: in the first one (sect.~\ref{s4.1}) we shall state the conjecture
 and discuss a few basic results (in particular on its finite temperature
 version) which will be needed in the following. In the second part 
 (sect.~\ref{s4.2} and \ref{s4.3})
 we shall review those results which are relevant
  for a comparison with the lattice. In particular, in sect.~\ref{s4.2}
  we shall only deal with
  the finite temperature (i.e. non-supersymmetric) realization of the
  correspondence  but, in this restricted field,  we shall try to
  keep our review as complete as possible. In sect.~\ref{s4.3} 
we shall mention  a few
 results concerning the supersymmetric theory (i.e. the zero temperature case)
 which, due to their generality,
 could be (despite the presence of supersymmetry) of some importance 
for the comparison that we are discussing.

\subsection{The AdS/CFT correspondence.}
\label{s4.1}
It is very important to stress that in going from string theory to QCD 
along the lines that we are discussing now, two distinct steps are 
needed.

The first one is the AdS/CFT correspondence, based on the 
 Maldacena conjecture~\cite{m98}, and further specified in the works of
 Witten~\cite{wn1} and Gubser, Klebanov and 
Polyakov~\cite{gkpn1}~\footnote{In particular in~\cite{wn1,gkpn1} 
the precise relation between the supergravity effective
action on one side and the correlation functions of the CFT on the other side
 was formulated for the first time. Both the results of~\cite{m98} and those
 of~\cite{wn1,gkpn1} are based on a set of
 earlier works~\cite{p97,p98,k97,gkt97,dps97}.}.
 This correspondence relates string 
theories on suitably chosen AdS manifolds with  conformally invariant field
theories whose symmetries depend on the internal manifold.

 The second step is the breaking of conformal invariance and (if present)
of supersymmetry, in order to obtain a candidate for a QCD-like theory.
In these lectures we shall follow 
the proposal of Witten~\cite{w98}, in which a QCD-like theory is obtained
 by compactifying the original theory with suitable 
boundary conditions. This proposal has several appealing features and originated
a large amount of papers, but it is not the unique possible choice. We shall
briefly comment on this further freedom below.

Let us now discuss in more detail these two steps.

\begin{itemize}
\item
The Maldacena conjecture relates the $M$ theory (or, depending on the case,
one of its superstring limits) in the $AdS_{d+1}\times{\bf
X}$ background 
to the large $N$ limit of a $d$ dimensional  conformal field theory.
 ${\bf X}$ is an  Einstein manifold whose particular form
 depends on the type of
field theory which we are interested to describe. In particular, by suitably
 choosing
${\bf X}$ it is possible to induce the presence
 on the field theoretic side of the
correspondence of supersymmetry, or of a $SU(N)$ gauge symmetry.
It is important to stress that, independently from the choice
of the background, the resulting field theory
 is always conformally invariant (this explains the abbreviation 
AdS/CFT which is used to denote this correspondence) and, as a consequence,
 is not confining.

In the following we shall only discuss this correspondence in two cases
\begin{description}
\item{1]}
In the first case we choose on the string side a type IIB 
superstring in the $AdS_5\times{\bf S^5}$ background. The corresponding field
theory turns out to be 
 the large $N$ limit of the $SU(N)$ ${\cal N} =4$ 
supersymmetric gauge theory in 4 dimensions. This will be the starting point to
obtain, following Witten's suggestion a candidate for a three dimensional
non-supersymmetric YM theory

\item{2]}
In the second case we choose to study, on the string side,
 $M$ theory on a  
$AdS_7\times{\bf S^4}$ background.
 This theory is mapped by
the Maldacena conjecture to the large $N$ limit of a   a six dimensional
$SU(N)$ type $(2,0)$ theory which is again supersymmetric and
 conformally  invariant. By compactifying the theory in two directions according
 to Witten's proposal we shall then obtain a candidate 
for a four dimensional
non-supersymmetric YM theory

\end{description}

It is important to stress that this AdS/CFT correspondence
  is formally only a conjecture. As a matter of fact
 one can recognize 
three levels of this conjecture. Let us discuss 
them in the most studied example of
type IIB string in the $AdS_5\times{\bf S^5}$ 
background (case 1 in the above list)

\begin{description}
\item{a]} In this case the ``weak'' statement
 is that the correspondence only holds between
supergravity on  $AdS_5\times{\bf S^5}$ 
 and the {\sl strong coupling limit} of large $N$ $SU(N)$
 ${\cal N} =4$ 
supersymmetric gauge theory in 4 dimensions
(this is the ``supergravity limit'' that we shall discuss below).
This level of the conjecture  has
 obtained by now so many confirmations 
(see for instance~\cite{rev_agmoo} for a thorough discussion of all
these checks) that it is commonly accepted as a firmly established result.

\item{b]} The ``normal'' level of the conjecture is the one that we have stated
at the beginning of this section. It extends the relation from  the supergravity
limit to the whole
type IIB 
superstring in the $AdS_5\times{\bf S^5}$ background, which is related with
 the large $N$ limit of the $SU(N)$ ${\cal N} =4$, with no constraint on the
 gauge coupling. This means that, with respect to the ``weak'' interpretation
 we are now pushing the correspondence {\sl outside the strong coupling limit}.
 If we want to to obtain from the AdS/CFT correspondence a QCD-like theory in
 the weak coupling limit (which is the ultimate goal of these lectures), 
 we must at least invoke this level of the conjecture. This is usually
 implicitly assumed in most of the papers that we shall discuss below.
 However essentially no check exists of the Maldacena conjecture at this level.

\item{c]} The ``strong'' level consists in assuming that the conjecture holds
also for the string theory at an arbitrary order in the loop expansion. From
the field theory side this would imply that we have informations not only in
the $N\to\infty$ limit but to any order in the $1/N$ expansion. As we have seen
in sect.~\ref{s3}, this is actually not so important since we have by now a
rather good control of the large $N$ limit on the Lattice side.

\end{description}

It is worthwhile to stress that 
(independently from the possible applications to QCD) 
the AdS/CFT correspondence is of 
great theoretical interest in itself. In some sense it represents the first 
nontrivial case in which we have been able to find the master field 
solution of a $d>2$ dimensional gauge theory in the large $N$ limit. 

At the same time it is the
first explicit description of a $d>2$ gauge theory in terms of a string theory.
 It seems somehow 
paradoxical that this remarkable result has been obtained for the first 
time in the case of a gauge theory which is not confining while the string
description of gauge theories has been based, from the very beginning, on the
intuitive picture of a string configuration spanning the minimal area of a
confining Wilson loop.
 The mechanism behind this apparent contradiction is very
instructive. The intuitive picture is indeed correct and also in the present
case the string is spanning the minimal area of the Wilson loop. However due to
the peculiar properties of the $AdS$ space the world-sheet of the string, in
order to minimize its area must wander deep 
 into the extra dimensions. This destroys the linear confining potential and
 leads to an effective $1/R$ behaviour.

\item
If we aim to reach a description of real QCD-like theories
 it is mandatory to break
the conformal invariance discussed above so as to recover a well behaved
confining potential. At the same time (if needed) we must somehow break the 
supersymmetry of the theory~\footnote{In principle the
breaking of supersymmetry is not a compelling requirement, since various
proposals exist for confining supersymmetric theories which are good candidates
to describe the phenomenology of strong interactions. However in these lectures
we shall follow a conservative attitude and look for non-supersymmetric
candidates for QCD. This is almost mandatory if we want to compare the results
with those obtained on the lattice where it is very difficult to implement
supersymmetry (for a recent discussion of
 this very delicate issue see for instance ref.~\cite{mont99}).}.
There are several possible ways to obtain these two results and we refer
to~\cite{rev_agmoo} and~\cite{rev_pdv} to a discussion of these options. In the
rest of these lectures we shall concentrate on the proposal suggested by Witten
in~\cite{w98}. The main appealing feature of this proposal is its simplicity, 
however one must always keep in mind that it is not the unique possibility. In
principle 
some of the problems that we shall discuss below and that seem to make
impossible a successful comparison with standard YM theories could be avoided
following other routes. 

Following~\cite{w98}, we can break the conformal invariance of the theory
 by  compactifying the theory in one (or more)
 direction(s). 
The presence of a new scale (the compactification radius) in the problem
 automatically breaks conformal invariance. If we then choose antiperiodic
 boundary conditions for the fermions in (one of) the compactified directions
 we also break supersymmetry. In fact as a consequence of the antiperiodic 
boundary conditions
the gauginos and and the adjoint scalars acquire a nonzero mass.
The important point in all these steps
 is that the Maldacena conjecture can be extended also to the
compactified version of the theory thus allowing to have an insight
in the infrared regime of the resulting gauge theory also in this case.
It is also important to stress, so as to avoid confusion, that the theory
obtained in this way is good candidate for a {\sl pure Yang Mills} theory.
 The term QCD which is often used in the literature is from this point of view
 rather misleading.

\end{itemize}

Let us now study the two interesting examples of $YM_3$ and $YM_4$.
 We choose 
to study first the case of 3d $YM$ which, for some technical reason 
turns out to be simpler, we shall later generalize the results to the 
more interesting case of 4d $YM$.

\subsubsection{The simplest example: $YM_3$}
\label{s4.1.1}

In this case we must start by studying type IIB 
superstring in the $AdS_5\times{\bf S^5}$ background. 
The Maldacena conjecture allows then to relate this theory  with
 the large $N$ limit of the $SU(N)$ ${\cal N} =4$ 
supersymmetric gauge theory in 4 dimensions. The pattern suggested by Witten
 to
break conformal invariance and supersymmetry is very simple in this case.
Both these goals can be reached,
by compactifying only one direction. There is a nice physical interpretation of
this recipe.
 If we choose to compactify the manifold in the time direction  then Witten's
 proposal is equivalent to study the original system at a nonzero temperature
 $T$, proportional to the inverse of the
 compactification radius $R_0$. For this reason we shall often call in the
 following the original $SYM$ theories as $T=0$ theories and the
 non-supersymmetric compactified ones as $T>0$ theories.
 In the $R_0\to 0$ (hence $T\to\infty$)
 limit we then obtain a three dimensional effective theory 
which has several features in
common with large $N$ 3d $YM$ and could hopefully be identified (at least in some
limit) with it.

Few comments are in order at this point:
\begin{description}
\item{\bf Coupling constants.}

It is important to follow the coupling constant identifications that emerge
from the two above steps. Let us define the
 coupling constant of the ${\cal N} =4$ $SU(N)$ theory as
$g_{YM,~({\cal N} =4)}^{(4)}$.
The Maldacena conjecture tells us that
$(g_{YM,~({\cal N} =4)}^{(4)})^2$  is proportional to 
the string coupling $g_s$.
We have seen in sect.~\ref{s3.4}
 that the large $N$ limit must be taken keeping the 't Hooft coupling 
 $\lambda\equiv Ng^2_{YM}$ finite. In the present case  the 't
 Hooft coupling is
 $\lambda\equiv N(g_{YM,~({\cal N} =4)}^{(4)})^2$.
  The coupling constant of the three dimensional
compactified theory can be expressed in terms of the four dimensional one 
as follows (this is a standard result in finite temperature gauge theories):
\eq
N(g_{YM}^{(3)})^2=\frac{N(g_{YM,~({\cal N} =4)}^{(4)})^2}{R_0}
\label{rel1}
\en
Thus while the original coupling $N(g_{YM,~({\cal N} =4)}^{(4)})^2$
 was dimensionless the 
3d one has the dimensions of a mass, which completely agrees with the expected
behaviour of $YM_3$ (see sect.~\ref{s2.3}). In the rest of this review
 we shall adopt the following convention to distinguish among the various
't Hooft couplings. We shall denote with $\tilde\lambda_d$ the couplings which
refer to the original $T=0$ supersymmetric $YM$ theories, where $d$ refers to the
dimension of the theory, while we shall denote with $\lambda_d$ the coupling of
the compactified non-supersymmetric theories. In this last case $d$ will denote
the number of uncompactified dimensions. Thus in the present case:
\eq
\tilde\lambda_4\equiv N(g_{YM,~({\cal N} =4)}^{(4)})^2~~~, \hskip1cm
\lambda_3\equiv N(g_{YM}^{(3)})^2~~.
\en
So that eq.(\ref{rel1}) becomes
\eq
\lambda_3=\frac{\tilde\lambda_4}{R_0}
\en

\item{\bf The supergravity limit.}

A crucial point for the following discussion is that we are actually unable to
study the string theory on the AdS manifold in its full complexity. As a matter
of fact we are bound to study the so called supergravity limit, in which the
string excitations are negligible and the string theory
reduces to supergravity. From the  AdS/CFT correspondence one can see
 that this region corresponds to the $\tilde\lambda_4 >> 1$ regime of the 
${\cal N} =4$ $SU(N)$ theory. In this limit the AdS/CFT correspondence is rather
well understood: it essential amounts to a correspondence between 
 supergravity fields on one side and  local operators of the gauge theory on the
 other side. The problem is that in this limit we may only have informations on
 the strong coupling sector of the gauge theory. If we now move to the
 compactified theory we face exactly the same problem. By using supergravity we
 may only have informations on the strong coupling regime of the theory that we
 hope to identify with $YM_3$.

\item{\bf Kaluza-Klein states.}

Another relevant problem is represented by the fact that in compactifying the
$S_5$ part of the original ten dimensional supergravity on $AdS_5\times S_5$ 
 a lot of Kaluza-Klein (K-K in the following) states are
 generated. Most of them have no counterpart in ordinary $YM$ theories and are
 expected to decouple in the $\lambda_3\to0$ limit.
 However in the limit in which
 the calculations can be performed there is still no evidence of such a
 decoupling. 
We shall come back to this problem when dealing with the glueball
 spectrum below.

\item{\bf Beyond supergravity.}

We have already said (and will repeat several time in the following) that
in order to test in a reliable way the predictions obtained from the AdS/CFT
correspondence we would need to extend them to the weak coupling regime of the
theory. The quest for such an extension
 will appear in all the tests that we shall discuss below.
However for small values of $\tilde\lambda_4$ the background geometry develops
a singular behaviour and the supergravity approximation breaks down. In this
regime one has to
study the string theory on the AdS manifold in its full complexity. 
This means in particular that one should be able to study the string theory
with background Ramond-Ramond charge in a singular background geometry. 
Despite several efforts few progress have been made in this direction up to
now. However it is important to stress that this seems to be only a technical
and not a conceptual obstacle and that it is well possible that in future this 
barrier could be overcome.
\end{description}

\subsubsection{Extension to $YM_4$}
\label{s4.1.2}
One can follow a procedure similar to the one outlined above to obtain a
non-supersymmetric four dimensional gauge theory which could hopefully be 
in the same universality class of $YM_4$. This time one must start by looking
at the $M$ theory in the $AdS_7\times S_4$ background. This theory is mapped by
the Maldacena conjecture to the large $N$ limit of a   a six dimensional
$SU(N)$ type $(2,0)$ theory which is supersymmetric and
 conformally  invariant. By compactifying the theory in two directions we then
 reach the desired four dimensional 
$SU(N)$ gauge theory. As a consequence of the
 compactification both supersymmetry and conformal invariance are lost, as it
 happened in the three dimensional case. However this time the 
relationship between the six dimensional gauge coupling and the four dimensional
one is much more subtle.
 Let us see this correspondence in more detail.
Let us denote with $R_1$ and $R_2$ the two compactification radii. In order to
 obtain a
reduced theory with the same features of $YM_4$, supersymmetry must be
 broken only in one of the two directions, let us choose it to be the one with
radius $R_2$, while $R_1$ will be the radius of the supersymmetry preserving
circle. Then the four dimensional gauge coupling constant $g^{(4)}_{YM}$ is
given by:
\eq
(g^{(4)}_{YM})^2=\frac{R_1}{R_2}
\en
which is adimensional, as in $YM_4$.
In order to reach the four dimensional theory that we hope to identify with
$YM_4$ both the radii must be sent to zero, however their role is very
different. Since in the large $N$ limit we want to keep the 't Hooft coupling
$\lambda_4=N (g^{(4)}_{YM})^2$  
to be finite, $R_1$ must go to zero in the large $N$
limit much faster than $R_2$. The remaining scale $R_2$ plays the role of an
ultraviolet cutoff for the four dimensional theory. Thus the gauge coupling
$g^{(4)}_{YM}$ must be thought of as the bare coupling at distances of order
$R_2$, and all dimensional quantities must be measured in units of $R_2$.
Remarkably enough the
situation is exactly the same that we have in LGT, with $R_2$ playing the same
role of the lattice spacing in LGT. 

If we aim to identify the  theory that we have found with $YM_4$, we must
require that, in the  $R_2\to0$ limit, $\lambda_4$ scales as follows: 
\eq
\lambda_4 \to -\frac{b}{\log(\Lambda_{QCD}R_2)}
\label{x2}
\en
with $b$ a suitable constant dictated by the Callan Symanzik equation.

However, exactly as in the three dimensional case discussed above we are only
able to study the large $\lambda_4$ regime of the theory and any test of a
behaviour like that of eq.(\ref{x2})
 is well beyond our present control of theory. Again, we are bound to study
the strong coupling regime of the theory. 

There are at this point two possible options: the first one is to try to
infer the small $\lambda_4$ behaviour of theory from the strong coupling
informations that we have. The second is to try to extrapolate real $YM$ to the
strong coupling limit and then compare with our findings. In both these
approaches the comparison with the lattice results plays a crucial role.

\subsection{Review of the results for the non-supersymmetric theories}
\label{s4.2}

All the attempts which have been made up to now
to compare the results obtained in the framework of the 
finite temperature version of  AdS/CFT correspondence
 with $YM$ theories
 dealt with
essentially only two topics: the glueball spectrum and the string tension. 
The present status of these calculations  is rather controversial. 
While a substantial agreement on the general pattern of both the glueball
spectrum and the string tension has been achieved, some conflicting 
results still exist and the 
whole issue is still evolving. In particular, the agreement with the LGT 
results which was claimed at the beginning has been lost in the most recent 
analyses. For this reason we shall avoid to collect in a table a tentative 
list of mass values as we did when reviewing the LGT results. 
 Instead we shall devote the next two sections to a review of 
the various attempts and results together with the open problems. 
 We shall then conclude by listing the general features 
on which a consensus has been reached. Let us finally mention that in the
following we shall write the coupling dependence of 
all the dimensional quantities 
 as a function of $\lambda_4$ in the four dimensions case and as a function 
of  $\tilde\lambda_4$ in the three dimensional one. The reason of this
asymmetric choice is that both  $\lambda_4$ and  $\tilde\lambda_4$ are
adimensional, and this greatly simplifies the discussion of the results.

\subsubsection{Glueball spectrum.}
\label{s4.2.1}
In principle
 the calculation of the glueball spectrum, 
at least for the $0^{++}$ state, is rather simple. 
In the supergravity limit the $0^{++}$
 glueball is mapped by the Maldacena conjecture into the dilaton field of the
 corresponding supergravity description. Its mass is then obtained by solving
 the dilaton wave equation. This calculation was first 
performed in~\cite{COOT,nunes,
 z98} where the spectrum of the first excited states of the $0^{++}$ and 
$0^{--}$ states in $d=3$ and of the $0^{++}$ and 
$0^{-+}$ glueballs in $d=4$ was obtained. The result had the correct dependence on
the ultraviolet cutoff ($R_0$ and $R_2$ respectively) and showed no explicit
dependence on $\tilde\lambda_4$ or $\lambda_4$.

The numerical values of the lowest states turned out to be of order unity if
measured in units of $1/R_0$ (or $1/R_2$).
These values were then compared with LGT
results~\footnote{Since there was no possibility 
to set the mass scale in units of some
other physical quantity as in LGT (we shall discuss below the problems involved
in the use of the string tension as a reference scale),
 the authors actually compared
the ratios of higher mass glueballs with respect to the $0^{++}$ one.} and a
good agreement was claimed. However it later appeared that such a claim was
probably not justified. In~\cite{cm} the mass of the $2^{++}$ glueball (both 
in $3$ and in $4$ dimensions) was obtained and turned out to be degenerate with
the $0^{++}$ one, a result which certainly disagree with the LGT estimate
reported in tab.s~\ref{tab7} and \ref{gb5}. 

At the same it was realized in~\cite{bmt}
 that in the three dimensional case the $0^{++}$ state associated with the
 graviton (in the supergravity limit) has a mass smaller than the one associated
 with the dilaton and hence must be considered as the lowest glueball state. If
 the various glueball masses are measured in units of this new fundamental mass,
 then the quantitative agreement with the LGT spectrum is definitely lost. 
 However a ``qualitative'' agreement with lattice results is still present.
 In~\cite{bmt} the authors also evaluated the glueball state with quantum
 numbers $1^{-+}$ and it turns out that\footnote{Notice that the degeneration
 between the $2^{++}$ and the $0^{++}$ states
 found in~\cite{cm} and further confirmed in~\cite{bmt} refers to the 
``dilaton'' $0^{++}$ glueball and not to the ``graviton'' one.}.
\eq
m(0^{++})<m(2^{++})<m(1^{-+})
\en
which is exactly the same pattern which emerges in LGT.
Let us stress that this is a rather non-trivial result. As mentioned
in sect.~\ref{s3.7} 
the fact that the $2^{++}$ state has a mass lower than that of the
$1^{-+}$ one is a fingerprint of QCD. 

The main problem of all these calculations is the presence of the unwanted
 Kaluza-Klein states discussed above. It turns out that in the supergravity
 limit the masses of these states are of the same order of those of the
 glueballs. As mentioned above this is nothing else that another
 signature that we are actually looking at the strong coupling regime of the
 theory. The approach to the problem followed in the papers discussed above was
 to simply neglect these states assuming that they should eventually
 decouple if one would be able to reach the weak coupling limit.
However it was noticed in~\cite{ORT} that, at least in the first order in the
string corrections, such a decoupling is not evident and the masses of the
 K-K states remain of the same order of that of the true glueball states.

It was thus proposed to study some suitable deformation of the supergravity
theory which could eliminate right from the beginning these states. The idea is
that there is not  a unique
 realization of the strong coupling theory which we
hope to identify with $YM$, but a whole family of theories which depend
 on one or
more free parameters. Thus in principle
 we could tune these parameters so as make the theory as
similar as possible to the weak coupling one. This is in some sense the same
philosophy of the so called "improved actions" in LGT. In this framework, 
the elimination of the K-K modes is certainly a step in the right direction.

This program was pursued in~\cite{rus,CORT,RS,CRST} and more recently in
in~\cite{SUZ}, Unfortunately only part of the K-K spectrum could be eliminated
in this way. The glueball spectrum obtained in this framework depends in the
most general case on three free parameters in $d=3$~\cite{RS} and two parameters
in $d=4$~\cite{CRST}, however it turns out that the mass ratios are very stable
as a function of these parameters. This interesting phenomenon, which points
toward the presence of some kind of universal behaviour has been discussed
in~\cite{mina}. 

Let us conclude with a last, positive, observation.
If one were able to extrapolate these mass gap
 calculations up to the weak coupling
limit then one should observe 
the scaling behaviour of eq.(\ref{cs}). In~\cite{COOT} a
first step has been made in this direction, by looking at the first string
correction of the mass spectrum. The authors found that the corrections are
negative. This means that, for a fixed ultraviolet cutoff the masses decrease as
the 't Hooft coupling $\lambda_4$ is decreased, in agreement with the expected
behaviour of eq.(\ref{cs}).

Let us summarize the main results.
\begin{itemize}
\item
Even if there is no quantitative agreement with the lattice estimates,
the pattern of the glueball masses (at least in $d=3$) is
correctly reproduced.
\item
The leading string corrections to the masses have the correct sign.
\item
The (dual) supergravity description of $YM$ can be generalized so as to
incorporate two (or three) free parameters. The mass ratios show a negligible
dependence on these parameters
\end{itemize}

\subsubsection{String Tension.}
\label{s4.2.2}
As we discussed in the introduction to LGT (sect.~\ref{s3.1.1}) 
in order to estimate the
string tension one must be able to evaluate the expectation value of Wilson
loops of large size. This problem is completely different from the one discussed
in the previous section and requires different tools. A method
 to compute these
Wilson loops in Super Yang Mills theories via supergravity was suggested
in~\cite{mal2,RY}. These ideas were then applied to the compactified theories
in which we are interested in~\cite{BISY2,RTY} and (as already predicted
in~\cite{w98}) a confining, linearly rising potential was indeed found.
It is important at this point to recall the discussion made in 
sect.~\ref{s3.6.3}~.

The Wilson loops which are studied in~\cite{BISY2,RTY} (and also in all the
other papers that we shall discuss in this section)
are the equivalent of what we called in sect.~\ref{s3.6.3}
 ``spacelike'' Wilson loops.
As such they do not give informations on the potential of the original theory
(which in fact, as we know, is not confining) but only on the dimensionally
reduced one. Since the original theory is not confining we expect a transition
(or a smooth crossover) between the two behaviours as the compactification
radius is shrunk to zero (i.e. as the temperature is increased). 

Let us call
$L$ the size of the Wilson loop (i.e. the distance between the quark and the
antiquark), let us study first the case of the $d=4$ ${\cal N}=4$ theory 
compactified to three dimensions. We expect that a confining potential
 appears in the limit
$\frac{L}{R_0} >>0$, i.e. when the distance between the quarks is much larger
than the compactification radius. In the opposite limit $\frac{L}{R_0}<<0$
on the contrary we expect to recover the Coulomb like behaviour of the 
${\cal N}=4$ $SYM$ in $d=4$. Indeed this is exactly the behaviour which was found
in~\cite{BISY2,RTY}. In the large $L$ limit it is thus possible to extract the
string tension whose value turns out to be
\eq
\sigma=\sqrt{\pi \tilde\lambda_4}\frac{\pi}{R_0^2}
\label{sigma3}
\en

A similar analysis can be performed also in four dimensions, leading to the
following expression for the string tension:
\eq
\sigma=\frac{8\pi}{27}\frac{\lambda_4}{R_2^2}
\label{sigma4}
\en
Both eq.(\ref{sigma3}) and (\ref{sigma4}) show that the string tension
 has the correct dimensions of  $(mass)^2$,
 however its dependence on the coupling constant shows that there is 
 a serious problem in the whole calculation. Moreover it is rather puzzling the
 fact that there is no signature of a $1/L$
 type term which in LGT arises from the quantum fluctuations of the effective
 string. Let us discuss these two problems in more detail.

\vskip0.5cm
\noindent
{\bf The $\lambda$ dependence of $\sigma$}

Let us address this problem directly in the four dimensional case.
 We saw in the previous section that the lowest glueball masses 
 were of order unity if measured in units of $1/R_2$ and showed
 no dependence on
 $\lambda_4$. On the contrary the square root of
the string tension measured in units of $1/R_2$ is proportional to
 $\sqrt{\lambda_4}$ which must be much larger than unity
 in the supergravity limit, which in turn
 is the only regime in which we can trust this
 solution. This has two unwanted consequences. 

First of all it completely
 disagrees with what we expect from both $YM_3$ 
 and $YM_4$ in the continuum limit
 (see sect.~\ref{s3.7}) where the ratio $\sqrt{\sigma}/M(0^{++})$ is of order
 unity. 

Second, it is telling us that we are actually testing the QCD string at
 very short distances, much shorter than the compactification radius,
 i.e. in a regime in which in ordinary $YM$ theories we do not expect to 
observe a ``string-like'' behaviour which is instead a 
peculiar feature of the large
distance infrared regime of Wilson loops.

As stressed in~\cite{GrOl}
the fact that $M(0^{++})$ and $\sqrt{\sigma}$ are of the same
order of magnitude and have the same dependence on the bare coupling
constant  is an  unavoidable consequence of
 the existence of a string-like description for the theory in which the
 glueballs come from closed strings.  It is amazing that
 this property does not hold if we describe $YM$ theories in
 the framework of the AdS/CFT correspondence which is explicitly constructed 
 to obtain a string  description of $YM$ theories. 
As we have seen above, this can
 be interpreted as a consequence of the fact that with the AdS/CFT results
 we are actually probing the short range regime of the the Wilson loop and that
 the confining regime that we observe has little to do with the real large
 distance potential of the theory. Thus we may hope that also
 this problem will be solved when we shall be able to overcome the supergravity
 limit.  An interesting proposal in this direction has been recently
 suggested in~\cite{Ol99}. The main point is that a log term appears in the 
 the string tension if the corrections
 induced by the quantum fluctuations of the string~\cite{GrOl1} are taken into
 account (we shall discuss in detail these corrections below, 
 when dealing with the L\"uscher term). The string tension becomes:
\eq
\sigma=\frac{8\pi}{27}\frac{\lambda_4}{R_2^2}+\frac{4\pi}{R_2^2}\log(R_2^2\mu^2)
+O(1/\lambda_4)
\label{sigma4bis}
\en
where $\mu$ is an arbitrary scale which is introduced to regulate the sum over
the modes of the string fluctuations. In principle we may use this additional
term to eliminate the unwanted $\lambda_4$ dependence in $\sigma$ by suitably
choosing the dependence of $\lambda_4$ on the ultraviolet cutoff $R_2$.
If we want to have a string tension
\eq
\sigma=\frac{c^2}{R_2^2}
\en
with some fixed constant $c$, such that the ratio $c/M(0^{++})$ agrees with the
results from LGT we must impose:
\eq
\lambda_4=\frac{27c^2}{8\pi}-27\log(R_2\mu)
\label{ol}
\en
This behaviour is compatible with the other constraints on $R_2$ and $\lambda_4$
if $\mu<<\frac{1}{R_2}$. In this case the log term dominates over the constant
and we recover the large $\lambda_4$ regime in which the result of 
eq.(\ref{sigma4}) was obtained. It is important to stress that eq.(\ref{ol})
imposes a constraint on the behaviour of the coupling
 constant $\lambda_4$ as a function of the ultraviolet cutoff which is different
 from the one due to asymptotic freedom (see eq.(\ref{x2})). It can be
 shown~\cite{Ol99} that these two constraints have somehow a symmetric role.
They determine the behaviour of the coupling
 $\lambda_4$ as a function of the cutoff in the strong and in the weak coupling
 regime respectively.

\vskip0.5cm
\noindent
{\bf The L\"uscher term}

We have seen in sect.~\ref{s3.5}
 that the area law is only the first dominant term of
the potential and that besides it we expect an universal subleading correction:
the L\"uscher term, which is due to the quantum fluctuations of the string.
In the framework of the AdS calculations that we are discussing, in the
supergravity limit, there is no signature of such a term. This problem was
noticed and discussed in~\cite{GrOl}. 

Once again a possible solution to the
problem is that such a term could appear
if  higher order string-like corrections are taken into account.
This type of calculations are  
 very delicate since they require a careful
treatment of the boundary conditions for the Wilson loop
 which in the finite temperature case is a rather non-trivial issue (see the
 comments in this respect in~\cite{DGO}and~\cite{dgt00}).
  A tentative in this direction
 was performed in~\cite{GrOl1},\cite{FGT} and~\cite{KSSW}
 and a term with the
desired $\frac{1}{L}$ behaviour was found, but with the wrong
sign!~\footnote{In~\cite{Naik} with a different calculation,
 a L\"usher term with the correct sign, was found. However in~\cite{Naik},
 by mimicking the type of calculation that we discussed in the section on the
 effective string picture in LGT (see sect.~\ref{s3.5}), only
 the quantum fluctuations of the transverse degrees of freedom were taken into
 account.}.

In view of
what we were saying before, i.e. of the fact that we are actually probing the
short range regime of the Wilson loops, the lack of a L\"uscher term is not
surprising. The same would happen also in LGT, where the $1/L$ correction
manifests itself only at distances much larger than the ultraviolet cutoff. 

However the fact that an $1/L$ term is present, but with the wrong sign rises a
different problem, which on the contrary seems to be
 rather serious. We have seen at the end of sect.~\ref{s3.5.3}
 that a very general requirement for
 the potential is that it must be a concave function of the interquark 
distance~\cite{bachas}
 and a L\"uscher term with the wrong sign violates this
requirement~\cite{GrOl1}. This problem was studied in detail in~\cite{DoPe}
where the authors 
discuss which conditions must be imposed on a AdS type theory  
so as to fulfill the concavity requirement in the induced gauge theory.

\subsection{A few results on the supersymmetric case.}
\label{s4.3}

{\bf String fluctuations in $AdS_5\times S^5$.}
As we mentioned above,  the calculations
 on the L\"uscher term discussed in the previous 
section are particularly delicate since they require a careful
treatment of the boundary conditions for the Wilson loop.
 Recently some progress has been made in this 
respect in the zero temperature case. In particular in~\cite{dgt00} a careful
discussion of the semiclassical fluctuations of strings in $AdS_5\times S^5$
based on the Green-Schwarz formalism can be found. In this paper the authors
also study, among other examples, the string corrections to the expectation
value of the Wilson loop in the ${\cal N}=4$ Super Yang Mills theory in $d=4$.

{\bf Loop equations.}

We have seen in sect.~\ref{s3.4}
 that a basic feature of the large N limit of $SU(N)$
 gauge theories is the fact that they satisfy the so called loop equations.
 The solution of these loop equations
 would be the sought for master field of the
 theory. Since these equations hold also in the strong coupling phase of the
 theory they are a perfect testground for the validity of the conjecture. This
 program was recently addressed in~\cite{DGO} where the
 zero temperature case i.e. the original ${\cal N}=4$ theory was 
studied\footnote{The loop equations may
 be derived in a rigorous form only  in the
 framework of the lattice discretization of the theory. Since 
for the moment there is no satisfactory formulation of
 supersymmetric theories on the lattice strictly speaking we cannot be sure that
 the loop equations still hold for these theories. 
However they can be derived, at least  formally also
 in the continuum theory, and in this case they can be extended also to the
 supersymmetric case.}. The loop
 equations have been indeed shown to hold, at least in
 all the cases studied by the authors. 

\noindent
{\bf Interaction between Wilson loops.}

Let us finally mention that some interesting results have also been obtained in
the study of the interaction between Wilson
loops~\cite{Zar,ESZ,SL}. These studies could offer new  
possibilities
of comparison with LGT where similar studies have been also performed.

\newpage

\section{Comparison between LGT and AdS/CFT results.}
\label{s5}
We have seen in the previous sections that $YM$ theories regularized 
on the lattice
have several features in common with the $YM$-like theories which one obtains in
the framework of the AdS/CFT correspondence. 

Let us summarize the results of
such comparison. For each of the following items we shall first compare 
the AdS/CFT results with LGT in the
 continuum limit, then we shall also mention the
difference between strong and weak coupling  LGT.
\begin{itemize}
\item {\bf Glueball spectrum.} 

Both theories have a mass gap. The qualitative features of the spectrum are the
same in the two theories, but there is no agreement at a quantitative level.
LGT calculations in the strong coupling limit disagree with the continuum
limit results (they also disagree with the AdS/CFT ones), but this disagreement 
becomes less significative as higher orders in the strong coupling expansion
are added. There is at least one example ($Z_2$ theory in three dimensions) in
which the expansion  can be pushed to so high order that the continuum limit
results are correctly recovered. 

\item {\bf String tension.} 

Both theories are confining, but the string tension which one obtains in the
AdS/CFT framework disagrees in many respects with the continuum limit LGT
results. First, the ratio $\sqrt{\sigma}/M(0^{++})$ has the wrong dependence on
the coupling constant. Second the L\"uscher term has (most probably) the wrong
sign. A similar situation also happens if one looks at strong coupling LGT. Also
 in this case  the ratio $\sqrt{\sigma}/M(0^{++})$ has the wrong scaling
behaviour. The L\"uscher term is exactly zero and it is well known that, as far
as the string tension is concerned, the
strong coupling phase is separated from the continuum limit by a phase
transition: the roughening transition.

\item {\bf Loop equations.} 

The loop equations hold both in strong coupling and weak coupling LGT. They can
be defined, at least formally, also in the  supersymmetric theories which appear
in the AdS/CFT correspondence. A preliminary analysis shows that they hold also
in this case.

\item {\bf String picture.}

The string description which is at the basis of the AdS/CFT approach is very
different from the LGT effective string discussed in sect.~\ref{s3.5} .
While the first one  fluctuates in the complementary space the second one
originates by an attempt to describe the string fluctuations in the 
transverse dimensions of the physical space. However in principle
 it is possible that the
second one could 
emerge as a large scale effective description from the first one.
An example of such a behaviour in a different context is
 the 3D gauge Ising model, in which the effective
string discussed in sect.~\ref{s3.5}
 emerges at large scale from the dynamics of the
Peierls contours at the level of the lattice spacing~\cite{Ising_string}.
 These Peierls contours
(with a suitable choice of the 3D lattice) are self-avoiding surfaces of very
high genus which are probably described by some unknown string
theory~\cite{Pol_book}
 of which the model discussed in sect.~\ref{s3.5}
 is a large
scale effective description.

\item {\bf Phase transitions.} 

It is by now clear that in the phase diagram of both $SU(2)$ and $SU(3)$
 LGT in four dimensions  with the Wilson action there is no phase transition
 separating the strong coupling regime from the continuum limit. For larger (but
 finite) values of $N$ or for different actions some lines of phase 
 transitions may appear in the phase diagram, but they do not represent an
 obstruction to  reach the continuum limit. Particular observables can undergo
 phase transitions (like the roughening one for the Wilson loop) in which some
 other correlation length of the theory (in the case of the Wilson loop the
 inverse of the stiffness of the surface bordered by the loop)  goes to
 infinity without affecting the true correlation length of the theory (i.e. the
 inverse of the $0^{++}$ mass). In the large $N$ limit the original Eguchi-Kawai
 model shows a phase transition which can be avoided by introducing suitable
 "twists" in the  boundary conditions  thus obtained 
 the so called twisted Eguchi-Kawai model. The possible presence of phase
 transitions in the AdS/CFT approach is an important open problem, for which no
 result has been obtained up to now. The fact that the qualitative features of
 the glueball spectrum are correctly predicted by the theory may be considered
 as a hint  that 
 also in this case there is no phase transition which forbids to reach the
 weak coupling regime.

\end{itemize}

\subsection{Concluding remarks}
\label{s5.1}
Let us try to extract the relevant outcomes of the above analysis.

It is clear that we may think of the AdS/CFT
approach as a new non-perturbative regularization, alternative to the
 lattice, of  $YM$ theories, in which the compactification radius in the extra
 dimensions plays the role of the lattice spacing $a$ in the lattice
 regularization. What is new with respect to the lattice approach is that in the
 AdS/CFT approach the ultraviolet cutoff, unlike the lattice spacing,
does not destroy the Lorentz symmetry of the theory. On top of this, if we study
the theory at the scale of the cutoff we see a higher dimensional theory, with
a much larger symmetry group and a clear string interpretation. What is missing
with respect to the lattice is that we lack a method to get rid of the
ultraviolet cutoff and reach the weak coupling limit. This would
require a better understanding of string theory on  AdS manifolds, 
and seems for the moment a too difficult task. Any progress in
this direction would play in the AdS/CFT context the same role which was played
by Montecarlo simulations in LGT.

Indeed the present status of the AdS/CFT 
physics strongly resembles the first years of
LGT, before the advent of Montecarlo simulations, when one was not even sure
that the continuum limit could be reached without finding some phase transition
in between, which could destroy all the nice properties found in the strong
coupling limit. The danger of the possible existence of such a phase
transition in the AdS/CFT approach has been stressed in~\cite{GO}.

Notwithstanding this analogies strong
coupling LGT and strong coupling AdS/CFT theory show very different behaviour.
This had to be expected since
the fixed point, which (thanks to universality) would justify a common
behaviour, is too far away. 
Thus it is meaningless to try to compare the two
strong coupling regimes. It is much better to compare the AdS/CFT results
directly with the weak coupling limit of LGT and see which  of the various
predictions seems to be less affected by the presence of the cutoff. In this
respect the qualitative agreement of the glueball spectrum as a function of the
angular momentum is a remarkable result. On the contrary, it seems that all the
physics concerning the Wilson loop and the string tension is, at least at the
present status of the analysis, definitely different from the weak coupling
expectations. 

However one should not care too much of these difficulties, because the goal is
certainly worthwhile. As a matter of fact the advent of MC simulations in the 
lattice community, besides the obvious advantages,
 also had the serious drawback that people felt less urged to 
reach a theoretical understanding of the nonperturbative physics of LGT. In
these last years progress in this direction has been much less
significative than twenty years ago. In this respect the AdS/CFT approach
represents a new fascinating idea and could help also people working in other
areas to have a fresh look to old problems.

\vskip 1cm
\centerline{\bf Acknowledgements}
\vskip 0.5cm\noindent
I am deeply indebted with M.~Bianchi, M.~Hasenbusch, I.~Pesando, 
 P.~Provero, A.~Zaffaroni and K.~Zarembo for several useful suggestions and a
 careful reading of an earlier version of these notes. 
I would also like to thank M.~Bill\'o, A.~D'Adda, 
F.~Gliozzi, U.~Magnea, K.~Pinn and S.Vinti with whom I wrote
  some of the works discussed in these lectures.
G.~Bali and C.~Morningstar are acknowledged for granting permission to
reproduce their figures.
This work was partially supported by the 
European Commission TMR programme ERBFMRX-CT96-0045.

\newpage
\appendix{}
\label{hkscapp}
\vskip 0.5cm
\section{{{Exercise Solutions.}}}

\subsection{Exercise 1: discuss some possible generalizations of the lattice
discretization of $SU(N)$ $YM$ theories.}
\renewcommand{\theequation}{E1.\arabic{equation}}
\setcounter{equation}{0}

The Wilson action can be generalized in three main directions (which may
obviously be combined together):
\begin{description}
\item{\bf Different representations.}

In this class of generalizations the basic variable, i.e.
 the group element on the
elementary plaquette, is unchanged, but we change the trace to a more general
real function on the group. We know from group theory that if we require
invariance  under the gauge transformation of eq.(\ref{gaugetransf})
 the most general function must be a linear combination of the group characters
(see exercise 3 for the definitions). Hence the most general for for the action
is
\eq
S=\sum_p\sum_r c_r Re\chi_r(U_p)
\en
where $U_p$ is a shorthand notation to denote the gauge variable of the $p$
plaquette and the $c_r$ are generalized coupling constants associated to the
various possible representations. 

Notice that at this level of generality we add no further complexity if we
expand in the character basis the Boltzmann factor, thus the action $S$  is 
often presented as
\eq
e^S=\prod_p\sum_r t_r \chi_r(U_p)
\en
where the $t_r$ are in general complicated functions of the couplings $c_r$,
but their explicit form is irrelevant and in this formulation they are usually
taken as the free parameters of the theory.

Among all the possible choices of $t_r$ a very
interesting one is the so called ``heat kernel'' action in which 
all the $t_r$ depend on a single parameter  $\beta$  as follows
\eq
t_r=d_re^{-\frac{C^{(2)}_r}{\beta N}} 
\en
where $d_r$ is the dimension of the representation and $C^{(2)}_r$
is the quadratic Casimir invariant for the representation $r$.
Apart from several interesting mathematical properties of this action, its
major reason of interest is that 
for any gauge group $SU(N)$, in the large $\beta$ limit the coefficients $D_r$
introduced in eq.(\ref{dierre})
become equivalent to the heat kernel ones: 
\eq
\lim_{\beta\to\infty} D_r(\beta)={\rm e}^{-\frac{C_r}{2N\beta}}~,
\label{242}
\en

\item{\bf Extended plaquette actions.}

The Wilson action can also 
be generalized by using loops larger than the elementary
plaquette. To each term we may associate a coupling constant. Its reason of
interest is that by suitably choosing these coupling constant one can improve
the scaling behaviour of the action.

\item{\bf Different lattices.}

Another obvious generalization is that of using different lattices, they can be
both random lattices or regular lattices with different elementary Brillouin
cells. In this last case the rotation symmetry is broken to subgroups different
from the cubic one and, again, this can help to keep under better control the
lattice artifacts.
\end{description}

\newpage
\subsection{Exercise 2: group theoretical analysis of the 
glueball states for the $SU(2)$ LGT in $d=3$.}
\renewcommand{\theequation}{E2.\arabic{equation}}
\setcounter{equation}{0}

The various glueball masses are labelled by their angular momentum.
Thus, in order to distinguish the various states of the spectrum one must 
construct operators with well defined angular momentum with respect of the
two-dimensional rotations group (remind that we are interested in spacelike
loops).  
Since we are working on a cubic lattice, 
where only rotations of multiples of 
$\pi/2$ are allowed, we must study the symmetry
properties of our operators with respect to a finite subgroup of the
two-dimensional rotations. 
Let us first ignore the effect of the lattice
discreteness and deal with the peculiar features which, already in the
continuum formulation,  the (2+1)  $SU(2)$ spectrum has with respect to 
the (3+1) dimensional $SU(3)$ spectrum.

\begin{description}
\item{a]} For  the $SU(2)$ model, we cannot define a charge
conjugation operator. The glueball states are thus labelled only by their 
angular momentum $J$ and by their parity eigenvalue $P=\pm$. 
The standard notation is $J^{P}$

\item{b]} In (2+1) dimensions it can be shown that all the states with 
angular momentum different from zero are degenerate in parity. Namely $J^+$ and
$J^-$ (with $J\neq 0$) must have the same mass.

\end{description}

\begin{figure}[t]
\centerline{\epsfxsize=10truecm\epsffile{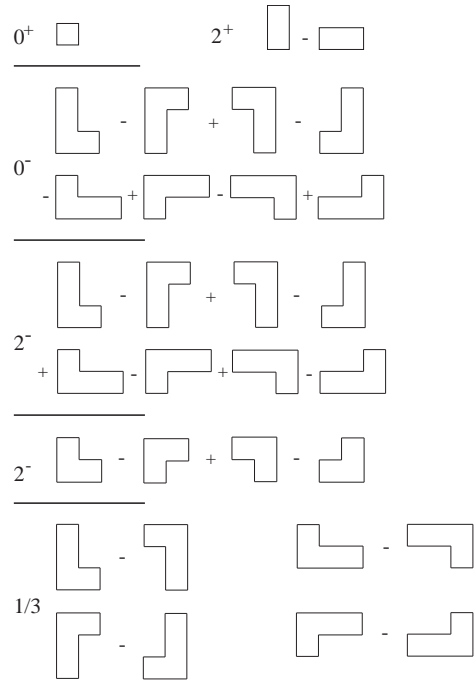}}
\caption{Some lattice realizations of the operators discussed in exercise~2. 
To clarify the role of the various symmetries, for the $2^-$ and 1/3 channels
we have shown  respectively two and four different realizations. 
For the $0^+$, $0^-$ and $2^+$ only the simplest possible realizations are
shown.}
\label{ads_gb}
\end{figure}

On the cubic lattice the group of two dimensional rotations and reflections 
becomes the dihedral group $D_4$. 
This group is non abelian, has eight elements and five
irreducible representations. Four of these are one-dimensional irreps, the last
one has dimension two. The group structure is completely described by the
table of characters~\cite{ham} which we have reported in tab.~\ref{character} .

 \begin{table}[ht]
 \caption{ \sl Character table for the group $D_4$}
 \label{character}
  \begin{center}
   \begin{tabular}{|l|c|r|r|r|r|}
   \hline
  & ${\bf 1}$ & $C_4^2$ & $C_4(2)$ & $C_2(2)$ & $C_{2'}(2)$ \\
   \hline
 $A_1$ & $1$ & $1$ & $1$ & $1$ & $1$ \\
 $A_2$ & $1$ & $1$ & $1$ & $-1$ & $-1$ \\
 $B_1$ & $1$ & $1$ & $-1$ & $1$ & $-1$ \\
 $B_2$ & $1$ & $1$ & $-1$ & $-1$ & $1$ \\
 $E$ & $2$ & $-2$ & $0$ & $0$ & $0$ \\
   \hline
   \end{tabular}
  \end{center}
 \end{table}

In the top row of tab.~\ref{character}
 are listed  the invariant classes of the group, and 
in the first column the irreducible representations. We followed the
notations of~\cite{ham} to label classes and representation (with the exception
of the class containing the identity which we have denoted with ${\bf 1}$
instead of the usual $E$ to
avoid confusion with the two-dimensional representation). The entries of the
table allow to explicitly construct the various representations and hence also 
the lattice operators 
which we are looking for.
The relationship of these operators with the various glueball 
states immediately follows from the group structure. In particular one can show
that:

\begin{description}
\item{a]} Only operators with angular momentum $J~(mod(4))$ can be 
constructed.
This is a common feature of all cubic lattice regularizations. It means that 
glueball states which in the continuum have values of $J$ higher than 3
appear on the lattice as secondary states in the family of the 
corresponding $J~(mod(4))$ lattice operator.

\item{b]} 
The four one-dimensional irreps are in  correspondence with the
even $J$  states. More precisely:
$$  0^+ \to A_1,\hskip 1cm 0^- \to A_2,\hskip 1cm 
 2^+ \to B_1 ,\hskip 1cm
 2^- \to B_2$$

This means that the discreteness of the lattice splits the degeneracy between
$2^+$ and $2^-$ which we discussed above. The splitting between these two states
gives us a rough estimates of relevance of the breaking of the full rotational
group due to the lattice discretization. Precise Montecarlo data~\cite{teper1}
 have shown that
in the scaling region this splitting is
essentially zero within the errors, 
in agreement with our expectation that approaching
the continuum limit the full continuum symmetries should be recovered.
Notice however that this is a very non-trivial result since the operators
associated to $2^+$ and $2^-$ on the lattice turn out to be very different.

\item{c]}
All the odd parity states are grouped together in the two-dimensional
irreducible representation $E$. This means that we cannot distinguish among 
them on the basis of the lattice symmetries. We can conventionally assume, say,
that the $J=1$ states have a mass lower than the $J=3$ ones, and that 
the $J=3$ thus appear as secondaries in the $J=1$ family. 
 In agreement with the above discussion, if the
full rotational symmetry is recovered, we expect the states belonging to
this family to be 
degenerate in parity and thus the lowest mass states, which
are the ones that we can measure more precisely, to
 appear as a doublet. Also this
prediction agrees with the data of~\cite{teper1}.

\end{description}

The simplest lattice operators, constructed according to the character table,
are shown in fig.~\ref{ads_gb}.

\newpage
\subsection{Exercise 3: Character Expansion for the $SU(N)$ group.}
\renewcommand{\theequation}{E3.\arabic{equation}}
\setcounter{equation}{0}

In this exercise we shall give some basic informations on the character
expansion for  $SU(N)$ groups, and shall then expand, as an example
the Wilson action in the character basis 
(for further details see Ref.~\cite{dz}).
Notice that, even if we deal in particular with
 the SU$(N)$ groups,
 most of the results that we shall discuss can hold for any Lie
group $G$ and with minor modifications also for discrete groups.

The irreducible characters $\chi_r(U)$ are the traces of the irreducible 
representations (labelled by $r$)  of
the group. They form a complete orthonormal basis for the class functions  on 
the group. A function $f(U)$ on the group is called a ``class function'' if it
satisfies the relation:
\eq
f(U)=f(VUV^{\dagger}) \hskip 2cm \forall\, V\in {\rm SU}(N)~.
\en
In particular, the characters themselves are class functions.
The pure gauge action, eq. (\ref{w1}), is a class function.

The following orthogonality relations between characters hold:
\eq
\int dU\, \chi_r(U)~\chi^*_s(U)=\delta_{r,s}~,
\label{c2}
\en
\eq
\sum_{r} d_r \chi_r(U~V^{\dagger})= \delta(U,V)~,
\label{c3}
\en
where $dU$ denotes the Haar measure (normalized to unity) on $SU(N)$ and
$d_r$ denotes the dimension of the $r^{\rm th}$ representation.

Besides the above orthogonality relations there are two 
 two other  integration formulas of the characters which turn out to be very
 useful in the construction of SC expansions:
\eq
\int dU\, \chi_r(V_1 U)~\chi_s(U^{\dagger}V_2)=\delta_{r,s}\frac{\chi_r(V_1 
V_2)}{d_r}~;
\label{c4}
\en
\eq
\int dU\, \chi_r(UV_1U^{\dagger}V_2)=\frac{1}{d_r}\chi_r(V_1)\chi_r(V_2)~.
\label{c5}
\en
Any class function can be expanded in the basis of the characters:
\eq
f(U)=\sum_r \chi_r(U) f_r~,
\en
where the sum is over the set of all irreducible representations of the group,
and the coefficients $f_r$ are given by
\eq
f_r\equiv \int dU\, \chi^*_r(U) f(U)~.
\en
Let us construct now the character expansion 
for the Wilson action.

The Boltzmann factor associated to each plaquette in
the Wilson action is (see eq.(\ref{w1})) :
\eq
{\rm e}^{\frac{\beta}{N} {\rm Re} {\rm Tr} U_{\mu\nu}(n)} 
= \sum_r F_r(\beta) \chi_r(U_{\mu\nu}(n))~,
\label{chexp}
\en
Notice that a factor 2 has been eliminated in the Boltzmann weight with
respect to eq.(\ref{w1}) so as to
avoid a double
counting of the plaquettes.
The coefficients $F_r$ are given by:
\eq
F_r(\beta)\equiv
\int dU\, {\rm e}^{\frac{\beta}{N} {\rm Re} {\rm Tr} U}\chi^*_r(U)
= \sum_{n=-\infty}^{\infty} {\rm det}\,I_{r_j-j+i+n}(\frac{\beta}{N})~.
\label{e9}
\en
The $r_j$'s are a set of integers labelling the representation 
$r$ and they are constrained by: $r_1\geq\cdots\geq r_N=0$.
The indices $1\leq i,j \leq N$ label the entries of the $N\times N$ matrix of 
which the determinant is taken and
$I_n(\beta)$ denotes the modified Bessel function of order $n$.

As a consequence of the factor $d_r$ at the denominator in eq.s  (\ref{c4},
\ref{c5}) the relevant coefficients in the character expansion (\ref{chexp}),
namely the ones that will appear in the strong coupling expansions, are not the
$F_r$ themselves, but the following normalized coefficients:

\eq
\label{dierre}
D_r(\beta)=\frac{F_r(\beta)}{d_r F_0(\beta)}~.
\en

Let us see in more detail two examples: the case of $SU(2)$ and the large $N$
limit. 

\vskip0.4cm
\noindent
{\bf Character expansion for $SU(2)$}
\vskip0.1cm
We can parameterize the most general
 matrix $U$ belonging to $SU(2)$ by using the Pauli $\sigma$
matrices.
\eq
U=cos(\theta/2) + i\vec{\sigma}\vec{n} sin(\theta/2) ~~~~~~~~~~
(0\leq\theta< 4\pi)~~,
\en
where $\vec{n}$ is a three dimensional normalized
 vector.

The normalized Haar measure is in this case
\eq
DU=\sin^2(\theta/2)\frac{d\theta}{2\pi}\frac{d^2\vec{n}}{4\pi}~~~~.
\en

The irreducible representations are labeled by the angular momentum
 $j=0,\frac12,1,\cdots$ and have dimension $d_j=2j+1$.

The character of the $j^{th}$ irreducible representation is:
\eq
\chi_j(U)=\frac{\sin(j+\frac12)\theta}{\sin(\theta/2)}
\label{e11}
\en

The Wilson action is in this case
\eq
exp\{\frac12 \beta\chi_{\frac12}(U)\}\equiv
exp\{\beta\cos(\theta/2)\}
\label{e12}
\en
where $U$ is the plaquette variable.

If we insert eq.(\ref{e11}) and (\ref{e12}) in eq.(\ref{e9}) we immediately
recognize one of the integral representations of the modified Bessel functions.

Thus the
expansion of the Wilson action in the character basis is 
\eq
exp\{\frac12 \beta\chi_{\frac12}(U)\}=\sum_j2(2j+1)
\frac{I_{2j+1}(\beta)}{\beta}\chi_j(U)
\label{e13}
\en
From this one can immediately recover the
 expression for the normalized coefficients $D_j(\beta)$ 
\eq
D_j(\beta)=\frac{I_{2j+1}(\beta)}{I_1(\beta)}
\label{esu2}
\en
\vskip0.4cm
\noindent
{\bf Character expansion in the large $N$ limit}
\vskip0.1cm
It is easy to see that in the large $N$ limit we find a finite value for the
coefficients $D_r(\beta)$ only if we simultaneously take the $\beta\to\infty$
limit while keeping the $\beta/N$ ratio fixed
 (in agreement with the 't Hooft prescription). 
In this limit the coefficients $F_r(\beta)$
 turn out to have  a very simple form.
In particular in the region
$\beta/N< 1$ one finds: 
\eqa
F_0(\beta/N)&\sim & {\rm e}^{\left(\frac{\beta}{2}\right)^2}~, \nonumber \\
F_f(\beta/N)&\sim &\frac{\beta}{2}\, 
{\rm e}^{\left(\frac{\beta}{2}\right)^2}~, 
\nonumber
\ena
where the index $f$ denotes the 
fundamental representation (whose dimension is $N$).  
The above relations imply that in the large $N$ limit 
\eq
D_f(\beta/N)=\frac{\beta}{2N}~.
\label{elargeN}
\en
Similar simplified relations hold also for higher representations.

\newpage
\subsection{Exercise 4: Evaluate the first  order of the strong 
coupling expansion of the Wilson loop in $YM$ theories.}
\renewcommand{\theequation}{E4.\arabic{equation}}
\setcounter{equation}{0}

\begin{figure}[hbt]
\centerline{\epsfxsize=8truecm\epsffile{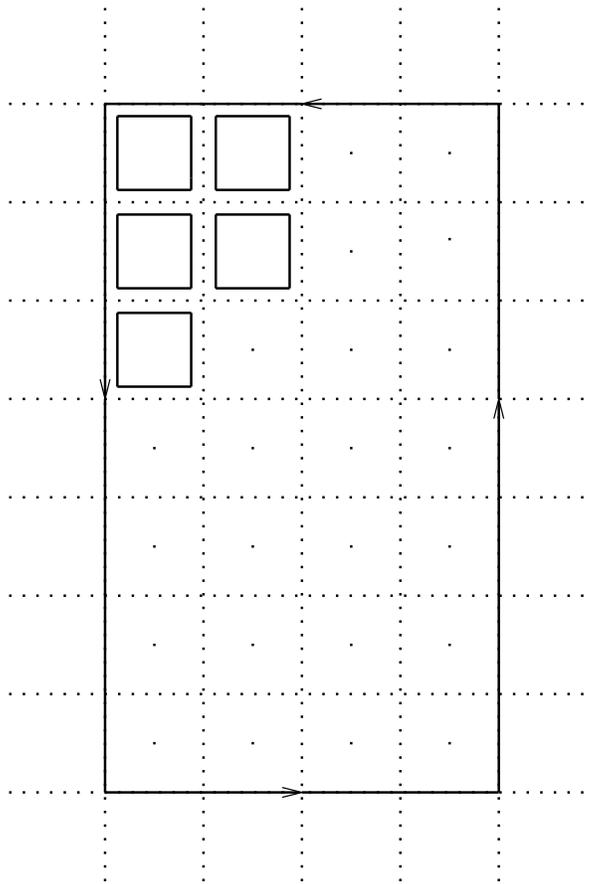}}
\caption{Strong coupling expansion for the Wilson loop.}
\label{ads_wilson2}
\end{figure}

Let us evaluate the first term in the strong coupling expansion 
the expectation value $\vev{\chi_f(U_c)}$, where $U_c$ is the ordered product of
gauge variables along a Wilson loop ``C'' of size $R\times T$ and $\chi_f$
denotes the character of the fundamental representation. 
Following eq.(\ref{vev}) the expectation value is defined as:
\eq
\vev{\chi_f(U_c)}=
\frac{\int \prod_{n,\mu} dU_\mu(n) \chi_f(U_c) e^{-S_W}}{Z}
\label{we.1}
\en
The first step is to insert in eq.(\ref{we.1}) the character expansion of the
Wilson action (see eq.(\ref{chexp})). 
The first non vanishing term in the
expansion is the one in which we keep for all the plaquettes
 inside the Wilson loop
(along the cristallographic plane, which ensures that we are keeping the
minimum number of terms) and only for them, exactly the term in the expansion
proportional to the fundamental representation. See fig.~(\ref{ads_wilson2}).
In this way for all the links inside the Wilson loop and along the border we
exactly find integrals of the type of eq.(\ref{c4}).

This allow to perform all
the group integrations in the expectation value,
link after link. 
Each plaquette inside the loop gives a contribution $F_f(\beta)$, they are
exactly $RT$. 
Each integration over the links gives a factor $1/d_f$. Again
these are $RT$ (one must take into account the fact
 that for each integration that we perform some of
the remaining links join together and thus at the end
the total number of link integrals is not $2RT$ but only $RT$). Finally 
we must keep into account the $Z$ factor at the denominator of the expectation
value. The simplest way to do this is to reorganize the strong coupling
expansion so as to factorize also in front of the numerator the same factor 
$Z$.
This simply amounts to normalize the coefficients of the expansion dividing them
by $F_0(\beta)$. Collecting everything together we find
\eq
\vev{\chi_f(U_c)}\sim \left(\frac{F_f(\beta)}{d_f F_0(\beta)}\right)^{RT}
\equiv D_f(\beta)^{RT}~~~.
\label{we.2}
\en
This explains, by the way, why we introduced the normalized coefficients
$D_r(\beta)$ in eq.(\ref{dierre}).

By using the definition of $\sigma$ (see eq.(\ref{sigdef}))
we immediately obtain from (\ref{we.2})
\eq
\sigma=-\log D_f(\beta)
\label{esig}
\en

\newpage

\newpage
\subsection{Exercise 5: Evaluate the first  order of the strong 
coupling expansion of the lowest glueball mass in $YM$ theories.}
\renewcommand{\theequation}{E5.\arabic{equation}}
\setcounter{equation}{0}

The solution of this exercise goes along the same lines of the one on the
Wilson loop. The only non trivial point is that we must find the surface of
minimal area bordered by the two plaquettes. If we are interested in the lowest
glueball state (i.e. the $0^{++}$ state) we know (see sect.~\ref{s3.1.2}) that
it is enough to study the connected correlator of two elementary spacelike
plaquettes (in the fundamental representation) located at two values of the
time coordinate $t_1$ and $t_2$ in the limit in which
$t\equiv|t_2-t_1|\to\infty$.
In this limit the connected correlator  decays
exponentially, i.e.
\eq
\vev{\chi_f(U_{ij}({\bf x},t_1))~\chi_f(U_{ij}({\bf x},t_2))} \sim e^{-Mt}
\en
where $(U_{ij}({\bf x},t_1))$ is the plaquette (with spacelike indices $(i,j)$)
located in the point $({\bf x},t_1)$ of the lattice, and $M$ is the mass of the
lowest glueball.
 It is easy to see that with this geometry the minimal surface connecting the
 two plaquettes is a long tube made of $4t$ plaquettes. Hence at the
 first order in the strong coupling expansion we have
\eq
\vev{\chi_f(U_{ij}({\bf x},t_1))~
\chi_f(U_{ij}({\bf x},t_2))} \sim D_f(\beta)^{4t}
\en
from which we immediately see that
\eq
M(0^{++})=-4\log(D_f(\beta))
\label{egb}
\en
\newpage
\subsection{Exercise 6: The effective string contribution to a
rectangular Wilson loop.}
\renewcommand{\theequation}{E6.\arabic{equation}}
\setcounter{equation}{0}

In this exercise we construct  the effective string theory
contribution for a Wilson loop in the infrared limit, assuming a
simple Nambu-Goto action for the string. 
As discussed in sect.~\ref{s3.5} the Nambu-Goto action 
reduces in this limit to the theory of 
$d-2$ free massless scalar fields. In this exercise we shall compute 
 the corresponding partition 
function $Z_{q}(R,T)$  which appears in
 eq.~(\ref{quantum}) following the discussion of ref.~\cite{ambj} (and references
 therein).

The Nambu string action is
given by the area of the 
world--sheet:
\be
S=\sigma\int_0^{T}d\tau\int_0^{R} d\varsigma\sqrt{g}\ \ ,\label{action}
\ee
where $g$ is the determinant of the two--dimensional metric induced on
the world--sheet by the embedding in $R^d$:
\ba
g=\det(g_{\alpha\beta})&=&\det\ \de_\alpha X^\mu\de_\beta X^\mu\ \ .\\
&&(\alpha,\beta=\tau,\varsigma,\ \mu=1,\dots,d)\nonumber
\ea
 and $\sigma$ is the string tension.
\par
The reparametrization and Weyl invariances of the action (\ref{action})
require a gauge choice for quantization. We choose the "physical gauge"
\ba
X^1&=&\tau\nonumber\\
X^2&=&\varsigma
\ea
so that $g$ is expressed as a function of the transverse degrees of freedom
only:
\ba
g&=&1+\de_\tau X^i\de_\tau X^i+\de_\varsigma X^i\de_\varsigma X^i\nonumber\\
&&\ \ \ +\de_\tau X^i\de_\tau X^i\de_\varsigma X^j\de_\varsigma X^j
-(\de_\tau X^i\de_\varsigma X^i)^2\\
&&\ \ \ \ (i=3,\dots,d)~.\nonumber
\ea
The fields $X^i(\tau,\varsigma)$ satisfy Dirichlet boundary conditions on $M$:
\be
X^i(0,\varsigma)=X^i(T,\varsigma)=X^i(\tau,0)=X^i(\tau,R)=0\ \ .
\ee
Due to the Weyl anomaly this gauge choice can be performed at the 
quantum level only in the
critical dimension $d=26$. However, the effect of
the anomaly is known to disappear at large distances \cite{olesen}, 
which is the region we are interested
in.\par
Expanding the square root in Eq.~(\ref{action}) we obtain, discarding 
terms of order $X^4$ and higher
\ba
S&=&\sigma R T+\frac{\sigma}{2}\int d^2\xi X^i (-\de^2) X^i\label{free}\\
\de^2&=&\de_\tau^2+\de_\varsigma^2~.
\ea
It is easy to see that this expansion of the action corresponds, for 
the partition function, to an expansion in powers of $(\sigma R T)^{-1}$.
Therefore the action (\ref{free}) describes the infrared limit of the 
model defined by Eq.~(\ref{action}), and will be relevant to the 
physics of large Wilson loops. 
The contribution of the fluctuations of the flux--tube to the Wilson 
loop expectation value in the infrared limit will be the partition 
function of our CFT, given by
\be
Z_{q}(R,T)\propto 
\left[\det(-\de^{2})\right]^{-\frac{d-2}{2}}~.\label{z1loop}
\ee
\vskip0.5cm\noindent
The determinant must be evaluated with Dirichlet boundary conditions. 
\par
The spectrum of $-\de^2$ with Dirichlet boundary conditions is 
given by the eigenvalues
\be
\lambda_{mn}=\pi^2\left(\frac{m^2}{T^2}+\frac{n^2}{R^2}\right)
\ee
corresponding to the normalized eigenfunctions
\be
\psi_{mn}(\xi)=\frac{2}{\sqrt{R T}}\sin\frac{m\pi\tau}{T}
\sin\frac{n\pi\varsigma}{R}~.
\ee
The determinant appearing in Eq.~(\ref{z1loop}) can be regularized with the 
$\zeta$-function technique: defining
\be
\zeta_{-\de^2}(s)\equiv\sum_{mn=1}^\infty\lambda_{mn}^{-s} \ \,\label{zeta}
\ee
the regularized determinant is defined through the analytic continuation
of $\zeta_{-\de^2}^\prime(s)$ to $s=0$:
\be
\det(-\de^2)=\exp\left[-\zeta_{-\de^2}^\prime(0)\right]\ \ .
\ee
\par
The series in Eq.~(\ref{zeta}) can be transformed, using the Poisson summation
formula, to read
\ba
&&\zeta_{-\de^2}(s)=-\frac{1}{2}\left(\frac{R^2}{\pi^2}\right)^s
\zeta_R(2s)+\frac{\sqrt{\pi}Im\tau\Gamma(s-1/2)}
{2\Gamma(s)}\left(\frac{R^2}{\pi^2}\right)^s\zeta_R(2s-1)\nonumber\\
&&\ \ \ +\frac{2\sqrt{\pi}}{\Gamma(s)}\left(\frac{T^2}{\pi^2}\right)^s
\sum_{n=1}^\infty\sum_{p=1}^\infty\left(\frac{\pi p}{n Im \tau}
\right)^{s-1/2}K_{s-1/2}(2\pi p n Im \tau)
\ea
where $\tau=iT/R$, $\zeta_R(s)$ is the Riemann $\zeta$ function 
and $K_\nu(x)$ is a modified Bessel function. The derivative 
$\zeta_{-\de^2}^\prime(s)$ can be analytically continued to $s=0$
where it is given by
\be
\zeta_{-\de^2}^{\prime}(0)=\log(\sqrt{2R})-\frac{i\pi\tau}{12}
-\sum_{n=1}^{\infty}\log(1-q^n)
\ee
where we have defined
\be
q\equiv e^{2\pi i\tau}~.
\ee
Introducing the Dedekind $\eta$-function
\be
\eta(\tau)=q^{1/24}\Pi_{n=1}^\infty(1-q^n)\label{etadef}
\ee
we obtain finally
\be
\det(-\de^2)=\exp[-\zeta_{-\de^2}^{\prime}(0)]=
\frac{\eta(\tau)}{\sqrt{2R}}
\ee
and
\be
Z_{q}(R,T)\propto\left[\frac{\eta(\tau)}{\sqrt{R}}
\right]^{-\frac{d-2}{2}}~.\label{zetaq}
\ee

\par
Substituting in Eq.~(\ref{quantum}) we obtain \cite{ambj}
\be
<W(R,T)>=e^{-\sigma RT+p(R+T)+k}\left[\frac{\eta(\tau)}{\sqrt{R}}
\right]^{-\frac{d-2}{2}}~.
\label{prediction}
\ee

Notice, as a concluding remark, that it is clear from the above discussion
that the Nambu-Goto action that we studied
in this exercise is only an instance
of a large 
class of bosonic effective string models 
  which reduce 
to the CFT studied in this exercise in the infrared limit. This is one of the
possible explanations for 
 the ``string universality'' discussed in sect.~\ref{s3.5.3}.

\vfill\eject


\begin{thebibliography}{99}


\bibitem{conj}
H.~B.~Nielsen and P.~Olesen,
{\sl ``Vortex Line Models For Dual Strings,''}
Nucl.\ Phys.\  {\bf B61} (1973) 45.


\bibitem{h74} 
G.~'t Hooft,
{\sl ``A Two-Dimensional Model For Mesons,''}
Nucl.\ Phys.\  {\bf B75} (1974) 461.

\bibitem{wilson}
K.~G.~Wilson,
{\sl ``Confinement Of Quarks,''}
Phys.\ Rev.\  {\bf D10} (1974) 2445.

\bibitem{polya}
A.~M.~Polyakov,
{\sl ``String Representations And Hidden Symmetries For Gauge Fields,''}
Phys.\ Lett.\  {\bf B82} (1979) 247.

A.~M.~Polyakov,
{\sl ``Gauge Fields As Rings Of Glue,''}
Nucl.\ Phys.\  {\bf B164} (1980) 171.

\bibitem{gene}
J.~Gervais and A.~Neveu,
{\sl``The Quantum Dual String Wave Functional In Yang-Mills Theories,''}
Phys.\ Lett.\  {\bf B80} (1979) 255.

\bibitem{nambu}
Y.~Nambu,
{\sl``QCD And The String Model,''}
Phys.\ Lett.\  {\bf B80} (1979) 372.


\bibitem{ek} 
T.~Eguchi and H.~Kawai,
{\sl``Reduction Of Dynamical Degrees Of Freedom In The Large N Gauge Theory,''}
Phys.\ Rev.\ Lett.\  {\bf 48} (1982) 1063.



\bibitem{m98} 
J.~Maldacena,
{\sl``The Large $N$ Limit of Superconformal Field Theories and
Supergravity,"}
Adv. Theor. Math. Phys. {\bf 2} (1998) 231.
hep-th/9711200.

\bibitem{w98}
E. Witten,
{\sl   ``Anti-de Sitter Space, Thermal Phase Transition, 
   And Confinement In Gauge Theories"}
   Adv.Theor.Math.Phys. {\bf 2}, 253 (1998),
   hep-th/9803131. 




\bibitem{os} 
K.~Osterwalder and R.~Schrader,
{\sl``Axioms For Euclidean Green's Functions,''}
Commun.\ Math.\ Phys.\  {\bf 31} (1973) 83.

K.~Osterwalder and R.~Schrader,
{\sl``Axioms For Euclidean Green's Functions. 2,''}
Commun.\ Math.\ Phys.\  {\bf 42} (1975) 281.


\bibitem{tj}
R.~Johnson and M.~Teper,
{\sl``Distinguishing J = 4 from J = 0 on a cubic lattice,''}
Nucl.\ Phys.\ Proc.\ Suppl.\  {\bf 73} (1999) 267
hep-lat/9808012.

\bibitem{arisue}
H.~Arisue and K.~Tabata,
{\sl``Strong coupling expansion of mass gap for Z(2) lattice gauge 
 theory in three dimensions,''}
Phys.\ Lett.\  {\bf B322} (1994) 224.


\bibitem{z2}
V.~Agostini, G.~Carlino, M.~Caselle and M.~Hasenbusch,
{\sl``The spectrum of the 2+1-dimensional gauge Ising model,''}
Nucl.\ Phys.\  {\bf B484} (1997) 331
hep-lat/9607029.

M.~Caselle and M.~Hasenbusch,
{\sl``Universal amplitude ratios in the 3D Ising model,''}
J.\ Phys.\ A {\bf A30} (1997) 4963
hep-lat/9701007.
%

\bibitem{master}
E. Witten, Proc. Nato Adv. Study Institute on Recent Developments in Gauge
Theories (Plenum Press, N.Y. 1980).


\bibitem{migmak}
Y.M.~Makeenko and A.A.~Migdal,
{\sl``Exact Equation for The Loop Average in Multicolor QCD,"}
Phys. Lett. {\bf 88B}, 135 (1979);

Y.~Makeenko and A.A.~Migdal,
{\sl``Quantum Chromodynamics as Dynamics of Loops,"}
Nucl. Phys. {\bf B188}, 269 (1981).


\bibitem{rev_loop}
A.A.~Migdal,
{\sl``Loop Equations and $1/N$ Expansion,"}
Phys. Rept. {\bf 102}, 199 (1984).

\bibitem{kk80}
V.A.~Kazakov and I.K.~Kostov,
{\sl``Nonlinear Strings in Two-Dimensional $U(\infty)$ Gauge Theory,"}
Nucl. Phys. {\bf B176}, 321 (1980)

\bibitem{das} S.M. Das, Rev. of Modern Physics {\bf 59} (1987) 235.

\bibitem{bcdp}
M.~Bill\'o, M.~Caselle, A.~D'Adda and S.~Panzeri,
{\sl``Finite temperature lattice {QCD} in the large N limit,''}
Int.\ J.\ Mod.\ Phys.\  {\bf A12} (1997) 1783
hep-th/9610144.




\bibitem{rough}
A.~Hasenfratz, E.~Hasenfratz and P.~Hasenfratz,
{\sl``Generalized Roughening Transition And Its Effect On The String Tension,
''}
Nucl.\ Phys.\  {\bf B180} (1981) 353.

C.~Itzykson, M.~E.~Peskin and J.~B.~Zuber,
{\sl``Roughening Of Wilson's Surface,''}
Phys.\ Lett.\  {\bf B95} (1980) 259.

\bibitem{lsw}
M.~L\"uscher, K.~Symanzik and P.~Weisz,
{\sl``Anomalies Of The Free Loop Wave Equation In The Wkb Approximation,''}
Nucl.\ Phys.\  {\bf B173} (1980) 365.


M. L\"uscher, 
{\sl``Symmetry Breaking Aspects Of The Roughening Transition In Gauge Theories,
''}
Nucl.\ Phys.\  {\bf B180} (1981) 317.

\bibitem{cfghpv}
M.~Caselle, R.~Fiore, F.~Gliozzi, M.~Hasenbusch, K.~Pinn and S.~Vinti,
{\sl``Rough interfaces beyond the Gaussian approximation,''}
Nucl.\ Phys.\  {\bf B432} (1994) 590
hep-lat/9407002.

\bibitem{lmw} 
M.~L\"uscher, G.~Munster and P.~Weisz,
{\sl``How Thick Are Chromoelectric Flux Tubes?,''}
Nucl.\ Phys.\  {\bf B180} (1981) 1.

\bibitem{width} 
M.~Caselle, F.~Gliozzi, U.~Magnea and S.~Vinti,
{\sl``Width of Long Colour Flux Tubes in Lattice Gauge Systems,''}
Nucl.\ Phys.\  {\bf B460} (1996) 397
hep-lat/9510019.

\bibitem{cfghp}
M.~Caselle, R.~Fiore, F.~Gliozzi, M.~Hasenbusch and P.~Provero,
{\sl``String effects in the Wilson loop: A high precision numerical test,''}
Nucl.\ Phys.\  {\bf B486} (1997) 245
hep-lat/9609041.


\bibitem{bachas}
C.~Bachas,
{\sl``Convexity of the Quarkonium Potential,''}
Phys.\ Rev.\ {\bf D33} (1986) 2723.





\bibitem{Bali:1992ab}
G.~S. Bali and K.~Schilling,
{\sl ``Static quark-antiquark potential: Scaling behaviour and finite size effects
 in SU(3) LGT,''}
\newblock Phys. Rev. {\bf D46}, 2636 (1992).

\bibitem{Bali:1997am}
G.~S. Bali, K.~Schilling, and A.~Wachter,
{\sl``Complete $O(v^2)$ corrections to the static interquark potential from 
 $SU(3)$ gauge theory,''}
\newblock Phys. Rev. {\bf D56}, 2566 (1997), hep-lat/9703019.



\bibitem{spacelike}
M.~Caselle, R.~Fiore, F.~Gliozzi, P.~Guaita and S.~Vinti,
{\sl``On the behavior of spatial Wilson loops in the high 
 temperature phase of LGT,''}
Nucl.\ Phys.\  {\bf B422} (1994) 397
hep-lat/9312056.



M.~Caselle and A.~D'Adda,
{\sl``The Spatial string tension in high temperature lattice gauge theories,''}
Nucl.\ Phys.\  {\bf B427} (1994) 273
hep-lat/9403027.


\bibitem{Morningstar:1999rf}
C.~J. Morningstar and M.~Peardon,
{\sl``The glueball spectrum from an anisotropic lattice study,''}
\newblock Phys. Rev. {\bf D60}, 034509 (1999), hep-lat/9901004.




\bibitem{wn1}
E.~Witten,
{\sl ``Anti-de Sitter space and holography,''}
Adv.\ Theor.\ Math.\ Phys.\  {\bf 2} (1998) 253
[hep-th/9802150].

\bibitem{gkpn1}
S.~S.~Gubser, I.~R.~Klebanov and A.~M.~Polyakov,
{\sl ``Gauge theory correlators from non-critical string theory,''}
Phys.\ Lett.\  {\bf B428} (1998) 105
[hep-th/9802109].

\bibitem{p98}
A.~M.~Polyakov,
{\sl ``String theory and quark confinement,''}
Nucl.\ Phys.\ Proc.\ Suppl.\  {\bf 68} (1998) 1
[hep-th/9711002].
%
\bibitem{p97}
A.~M.~Polyakov,
{\sl ``Confining strings,''}
Nucl.\ Phys.\  {\bf B486} (1997) 23
[hep-th/9607049].

\bibitem{k97}
I.~R.~Klebanov,
{\sl ``World-volume approach to absorption by non-dilatonic branes,''}
Nucl.\ Phys.\  {\bf B496} (1997) 231
[hep-th/9702076].

\bibitem{gkt97}
S.~S.~Gubser, I.~R.~Klebanov and A.~A.~Tseytlin,
{\sl ``String theory and classical absorption by three-branes,''}
Nucl.\ Phys.\  {\bf B499} (1997) 217
[hep-th/9703040].

\bibitem{dps97}
M.~Douglas, J.~Polchinski and A.~Strominger,
{\sl ``Probing five-dimensional black holes with D-branes,''}
JHEP {\bf 9712} (1997) 003
[hep-th/9703031].

\bibitem{mont99}
I.~Montvay,
{\sl ``Simulation of QCD and other similar theories,''}
hep-lat/9909020.


\bibitem{COOT}
C. Csaki, H. Ooguri, Y. Oz, J. Terning
{\sl  ``Glueball Mass Spectrum From Supergravity"}
  JHEP {\bf 9901}, 017 (1999),
  hep-th/9806201,

\bibitem{nunes} R.de Mello Koch, A. Jevicki, M. Mihailescu, J. P. Nunes, 
{\sl ``Evaluation of glueball masses from supergravity",}
Phys.Rev. D58 (1998) 105009, hep-th/9806125


\bibitem{z98}
M.Zyskin
{\sl ``A note on the glueball mass spectrum'',}
\PL{B439} (1998) 373, hep-th/9806128.


\bibitem{cm}
N.R.Constable and R.C Myers,
{\sl ``Spin-two glueballs, positive energy theorems and the AdS/CFT
correspondence",}
 JHEP 9910 (1999) 037, hep-th/9908175.

\bibitem{bmt}
R.Brower, S.D.Mathur and C.-I Tan,
{\sl ``Discrete spectrum of the graviton in the $AdS^5$ black hole background",}
hep-th/9908196.

\bibitem{ORT}
H. Ooguri, H.Robin, J. Tannenhauser, 
{\sl  ``Glueballs and their Kaluza-Klein cousins"}
  Phys.Lett. {\bf B437}, 77(1998),
  hep-th/9806171.

\bibitem{rus}
J. G. Russo,
{\sl  ``New Compactifications of Supergravities and Large N QCD"}
  Nucl.Phys. {\bf B543}, 183 (1999),
  hep-th/9808117.

\bibitem{CORT}
C. Csaki, Y. Oz, J. Russo, J. Terning,
{\sl  ``Large N QCD from Rotating Branes"}
  Phys.Rev. {\bf D59}, 065012 (1999),
  hep-th/9810186. 

\bibitem{RS}
J.G. Russo, K. Sfetsos, 
{\sl  ``Rotating D3-branes and QCD in three dimensions"}
  Adv. Theor. Math. Phys. {\bf 3}, 131 (1999),
  hep-th/9901056. 

\bibitem{CRST}
C. Csaki, J. Russo, K. Sfetsos, J.Terning,
{\sl  ``Supergravity Models for 3+1 Dimensional QCD"}
  Phys.Rev. {\bf D60}, 044001 (1999),
  hep-th/9902067.

\bibitem{SUZ}
  K. Suzuki,
{\sl  `` $D0 -D4$ systems and $QCD_{3+1}$"}
  hep-th/0001057.

\bibitem{mina}
J.A. Minahan, 
{\sl  ``Glueball mass spectra and other issues for
   supergravity duals of QCD models"}
  JHEP {\bf 9901}, 020 (1999),
  hep-th/9811156.

\bibitem{mal2}
J. Maldacena, 
{\sl   ``Wilson loops in large N field theories"}
   Phys.Rev.Lett. {\bf 80}, 4859 (1998),
   hep-th/9803002.

\bibitem{RY}
S.J. Rey, J. Yee, 
{\sl   ``Macroscopic strings as heavy quarks in large N
   gauge theory and Anti-de Sitter supergravity"}
   hep-th/9803001;

\bibitem{BISY2}
A. Brandhuber, N. Itzhaki, J. Sonnenschein, S. Yankielowicz,
{\sl  ``Wilson loops, confinement, and phase transitions in large N
   gauge theories from supergravity"}
  JHEP {\bf 9806}, 001 (1998),
  hep-th/9803263.

\bibitem{RTY}
S.J. Rey, S. Theisen,  J. Yee,  
{\sl   ``Wilson-Polyakov loop finite temperature in large N
   gauge theory and Anti-de Sitter supergravity"}
   Nucl.Phys. {\bf B527}, 171 (1998),
   hep-th/9803135.



\bibitem{DGO} Nadav Drukker, David J. Gross, Hirosi Ooguri,
{\sl ``Wilson Loops and  Minimal Surfaces",}
\PR{D60} (1999) 125006,  hep-th/9904191

\bibitem{dgt00} Nadav Drukker, David J. Gross, Arkady A. Tseytlin,
{\sl ``Green-Schwarz string in $ADS_5\times S^5$: Semiclassical partition
function,"}
  hep-th/0001204

\bibitem{GrOl}  J. Greensite, P. Olesen, 
{\sl ``Remarks on the Heavy Quark Potential in the Supergravity Approach"}
JHEP 9808 (1998) 009, hep-th/9806235

\bibitem{Ol99} P. Olesen, 
{\sl ``Some speculations on the gauge coupling in the ADS/CFT approach"}
JHEP 9904 (1999) 001, hep-th/9904034



\bibitem{GrOl1}J. Greensite, P. Olesen, 
{\sl ``Worldsheet Fluctuations and the Heavy Quark Potential 
  in the AdS/CFT Approach",}
JHEP 9904 (1999) 001, hep-th/9901057

\bibitem{FGT} S. Forste, D. Ghoshal, S. Theisen,
{\sl ``Stringy Corrections to the Wilson Loop in N=4 Super Yang-Mills Theory",}
 JHEP 9908 (1999) 013, hep-th/9903042.

\bibitem{KSSW} Y. Kinar, E. Schreiber, J. Sonnenschein and N. Weiss,
{\sl ``Quantum fluctuations of Wilson loops from string models",}
 hep-th/9911123.

\bibitem{Naik} S. Naik,
{\sl ``Improved heavy quark potential
 at finite temperature from anti-de Sitter supergravity",} 
\PL{B464} (1999) 73, hep-th/9904147.

\bibitem{DoPe}
H.Dorn and V.D.Pershin,
{\sl``Concavity of the $Q\bar Q$ potential in the ${\cal N}=4$ super 
 Yang-Mills gauge theory and AdS/CFT duality"}
hep-th/9906073
 
\bibitem{Zar}
K.~Zarembo,
{\sl``Wilson loop correlator in the AdS/CFT correspondence,''}
Phys.\ Lett.\  {\bf B459} (1999) 527
hep-th/9904149.

\bibitem{ESZ}
J.~Erickson, G.~W.~Semenoff and K.~Zarembo,
{\sl``BPS vs. non-BPS Wilson loops in N = 4 supersymmetric Yang-Mills theory,
''}
Phys.\ Lett.\  {\bf B466} (1999) 239
hep-th/9906211.

\bibitem{SL}  J. Sonnenschein and A. Loewy,
{\sl ``On the supergravity evaluation of wilson loops correlators in confining
 theories",}
 hep-th/9911172.

\bibitem{Ising_string}
M.~Caselle, F.~Gliozzi and U.~Magnea,
{\sl``Fluid interfaces in the 3-d Ising model as a dilute gas of handles,''}
Physica {\bf A215} (1995) 21.


\bibitem{Pol_book}
A.M.~Polyakov,
{\sl``Gauge Fields and Strings,"}
{\it  Harwood (1987) 301 p. (Contemporary
Concepts in Physics, 3)}.


\bibitem{GO}
D.J. Gross, H. Ooguri, 
{\sl  ``Aspects of large N gauge theory dynamics as seen by string theory"}
  Phys.Rev. {\bf D58}, 106002 (1998),
  hep-th/9805129. 


\bibitem{ham} M. Hamermesh, {\it Group Theory and its Application to Physical
Problems}, Reading Mass.: Addison-Wesley, 1962



\bibitem{ambj}
J.~Ambjorn, P.~Olesen and C.~Peterson,
{\sl``Three-Dimensional Lattice Gauge Theory And Strings,''}
Nucl.\ Phys.\  {\bf B244} (1984) 262.

\bibitem{olesen}
P.~Olesen,
{\sl``Strings And QCD,''}
Phys.\ Lett.\  {\bf B160} (1985) 144.

\vskip 0.4cm
{\Large{\bf Books and Reviews on L.G.T.}}

\bibitem{Rebbi} C. Rebbi, {\it Lattice Gauge Theories and Montecarlo
Simulations},
		 World Scientific 1983. 

\bibitem{Creutz83} M. Creutz, {\it Quarks, Gluons, and Lattices}, Cambridge 
		   University Press, 1983.
\bibitem{Creutz92} {\it Quantum Fields on the Computer}, Ed. M. Creutz, 
World Scientific, 1992.

\bibitem{MM} I. Montvay and G.~M\"unster, {\it Quantum Fields on a 
Lattice}, 
Cambridge University Press, 1994.

\bibitem{idbook} C. Itzykson and J.-M. Drouffe, {\it Statistical Field 
Theory}, 
Cambridge University Press, 1989.
(In particular chapters 6,7 and 8 are devoted to L.G.T.)

\bibitem{Kogut79RMP}
J.~B.~Kogut,
{\sl``An Introduction To Lattice Gauge Theory And Spin Systems,''}
Rev.\ Mod.\ Phys.\  {\bf 51} (1979) 659.

\bibitem{Kogut83RMP}
J.~B.~Kogut,
{\sl``A Review Of The Lattice Gauge Theory Approach To Quantum Chromodynamics,
''}
Rev.\ Mod.\ Phys.\  {\bf 55} (1983) 775.

\bibitem{dz} J.M. Drouffe and J.B. Zuber,
{\sl ``Strong Coupling and Mean Field Methods in Lattice Gauge Theories,''}
 Phys. Rep. {\bf 102}(1983) 1.


\bibitem{gupta} 
R.~Gupta,
{\it Introduction to lattice QCD.}

Les Houches lectures (1997 Summer School), hep-lat/9807028
\bibitem{luscher}
M. L\"uscher
{\it Advanced Lattice QCD}

Les Houches lectures (1997 Summer School), hep-lat/9802029

\bibitem{teper1} 
M.~J.~Teper,
{\it SU(N) gauge theories in (2+1) dimensions}
Phys.\ Rev.\  {\bf D59} (1999) 014512
hep-lat/9804008.

\bibitem{teper2}
M.~J.~Teper,
{\it Glueball masses and other physical properties of SU(N) gauge 
theories  in D = 3+1: A review of lattice results for theorists.}
hep-th/9812187.

\bibitem{bali}
G. S. Bali,
{\sl``QCD forces and heavy quark bound states,''}
hep-ph/0001312

\bibitem{latproc} LATTICE93 \NPBPS{34} (1994); 

			  LATTICE94 \NPBPS{42} (1995); 

			  LATTICE95 \NPBPS{47} (1996); 

			  LATTICE96 \NPBPS{53} (1997); 

			  LATTICE97 \NPBPS{63} (1998); 

			  LATTICE98 \NPBPS{73} (1999). 



\vskip 0.4cm
{\Large{\bf Reviews on the AdS/CFT correspondence.}}



\bibitem{rev_pdv}
P. Di Vecchia,  
{\sl   ``Large N  gauge theories and AdS/CFT correspondence",}
   hep-th/9908148.
\bibitem{rev_agmoo}
O.Aharony, S.S.Gubser, J.Maldacena, H.Ooguri and Y.Oz,  
{\sl   ``Large N  field theories, string theory and gravity",}
   hep-th/9905111.



\end{thebibliography}
\end{document}